\documentclass[twocolumn,showpacs,superscriptaddress,preprintnumbers,amsmath,amssymb,longbibliography]{revtex4-1}

\usepackage[bookmarks=true,colorlinks=true,citecolor=blue,linkcolor=blue,urlcolor=magenta]{hyperref}

\usepackage{dcolumn}
\usepackage{multirow}
\usepackage{dcolumn}
\usepackage{bm}
\usepackage{xfrac}
\usepackage{calc}
\usepackage{graphicx}
\usepackage[tight,footnotesize]{subfigure}
\usepackage{color}
\usepackage{longtable}
\usepackage{dashrule}
\usepackage{multirow}%
\usepackage{rotating}

\usepackage{float}
\usepackage{perpage}
\usepackage{morefloats}

\usepackage{latexsym}
\usepackage{gensymb} 
\usepackage{bbm}
\usepackage[dvipsnames]{xcolor}
\ifpdf
\graphicspath{
	{figures/}
	{figures/ActDensity/}
	{figures/QQpotential/}
	{figures/SelfInt/}	
	{figures/stringWidth/}		
	{figures/6updates_WidthFits/}
}
\else
\graphicspath{
	{figures/}
	{figures/ActDensity/}	
	{figures/QQpotential/}	
	{figures/SelfInt/}
	{figures/stringWidth/}	
	{figures/6updates_WidthFits/}
}
\fi

\DeclareFontFamily{OT1}{pzc}{}
\DeclareFontShape{OT1}{pzc}{m}{it}{<-> s * [0.900] pzcmi7t}{}
\DeclareMathAlphabet{\mathpzc}{OT1}{pzc}{m}{it}

\begin{document}
\preprint{AIP/123-QED}
\title[]{Quantum delocalization of strings with boundary action \\ in  Yang-Mills theory}
\author{A. S. Bakry}
\affiliation{Institute of Modern Physics, Chinese Academy of Sciences, Gansu 730000, China}
\author{M. A. Deliyergiyev}\email[]{maksym.deliyergiyev@ujk.edu.pl}
\affiliation{Institute of Physics, The Jan Kochanowski University in Kielce, 25-406, Poland}
\affiliation{Institute of Modern Physics, Chinese Academy of Sciences, Gansu 730000, China}
\author{A. A. Galal}
\affiliation{Department of Physics,  Al Azhar University, Cairo 11651, Egypt}
\author{\\A. M. Khalaf}
\affiliation{Department of Physics,  Al Azhar University, Cairo 11651, Egypt}
\author{M. KhalilA Williams}
\affiliation{Department of Physics,  Al Azhar University, Cairo 11651, Egypt}
\affiliation{Department of Mathematics, Bergische Universit\"at Wuppertal, 42097 Germany}
\affiliation{Department of Physics, University of Ferrara, Ferrara 44121, Italy}
\affiliation{Research and computing center, The Cyprus Institute, Nicosia 2121, Cyprus}

\date{January 8, 2020}

\begin{abstract}
  
The width of the quantum delocalization of the QCD strings is investigated in effective string models beyond free Nambu-Goto approximation. We consider two Lorentzian-invariant boundary-terms in the L\"uscher-Weisz string action in addition to self-interaction term equivalent to two loop order in the (NG) string action. The geometrical terms which realize the possible rigidity of the QCD string is scrutinized as well. We perform the numerical analysis on the 4-dim pure SU(3) Yang-Mills lattice gauge theory at two temperature scales near deconfinement point. The comparative study with this QCD string model targets the width of the energy profile of static quark-antiquark system for color source separation  $0.5 \le R \le 1.2$ fm. We find the inclusion of rigidity properties and symmetry effects of the boundary action into the string paradigm to reproduce a good match with the profile of the Mont-Carlo data of QCD flux-tube on this distance scale.
  
\end{abstract}
\pacs{12.38.Gc, 12.38.Lg, 12.38.Aw}
\keywords{QCD Phemenonlogy, Effective bosonic string, Nambu-Goto action, Polyakov-Kleinert action, Montecarlo methods, Lattice Guage Theory}
\maketitle

\section{Introduction}
  Understanding the confining force in elementary particle physics is essential for modeling the hadron structure. It remains, however, insurmountable to put an analytic form to the binding forces between quarks in the non-perturbative region of QCD from first principles.

  Perturbative QCD provides a good description to the short-distance aspects of the $Q\bar{Q}$ potential of a two-body (Coulombic) one-gluon-exchange (OGE) interaction potential~\cite{Brambilla:2013vx}. Discussions concerning the intermediate and large distances, however, are usually carried out either on phenomenological bases, making use of the strong-coupling expansion \cite{PhysRevD.11.395,Nakano:2003qx} or lattice simulations.

  Lattice simulations have shown that the QQ potential is linearly rising~\cite{Creutz:1980zw,Creutz:1983ev,Creutz:1980hb,Creutz:1980vq}. In a string phenomenology, the linear rise is consistent with the formation of a stringlike flux tube~\cite{FUKUGITA1983374, Cea:1993pi, Cea:2015wjd, FLOWER1985128, Otto:1984qr, AM, 10.1143, PhysRevD.36.3297, SOMMER1987673, etde_6934022, DIGIACOMO1990441, Bali:1994de} linking the color sources. The quantum fluctuations of the string~\cite{Luscherfr} produce the well-known Coulomb-like sub-leadingcorrections to the QQ potential, namely, the well-known Lüscher term.

  The effective description is expected to hold over distance scale  $1/T_c$~\cite{Caselle:2012rp} where the effects of the intrinsic thickness of the flux tube diminish. Many gauge models have accurately verfied the L\"uscher correction to the potential~\cite{Juge:2002br, HariDass2008273, Caselle:2016mqu,caselle-2002,Pennanen:1997qm,Brandt:2016xsp}.
  

    The width due to the quantum delocalizations of the string grows logarithmically~\cite{Luscher:1980iy} as the two color sources are pulled apart. The character of logarithmic broadening is expected for the baryonic junction~\cite{PhysRevD.79.025022} as well. The string model predicts, in addition, a logarithmic broadening~\cite{Luscher:1980iy} for the width profile of the string delocalization at very low temperatures. This also has been observed in several lattice simulations corresponding to the different gauge groups~\cite{Caselle:1995fh,Bonati:2011nt,HASENBUSCH1994124,Caselle:2006dv,Bringoltz:2008nd,Athenodorou:2008cj,Juge:2002br,HariDass:2006pq,Giudice:2006hw,Luscher:2004ib,Pepe:2010na,Bicudo:2017uyy}

    
    In the high temperature regime of QCD, the string's broadening becomes linear~\cite{allais,Gliozzi:2010zv,Caselle:2010zs,Gliozzi:2010jh, PhysRevD.82.094503,Bakry:2010sp,Bakry:2012eq}. However, the comparison with the lattice Monte-Carlo data substantial show deviations~\cite{PhysRevD.82.094503,Bakry:2010sp,Bakry:2011zz,Bakry:2012eq} from the free string behavior very close to the deconfinement point. The discrepancies take place at distances at which the leading-order string model predictions are valid~\cite{Luscher:2002qv} at zero temperature. In the baryon ~\cite{Bakry:2014gea,Bakry:2016uwt,Bakry:2011kga} similar short distances mismaches are found~\cite{refId0}.

   Several phenomena can involve the stringy character of the gluonic sector of $SU(3)$ Yang-Mills theory~\cite{Caselle:2015tza,GIDDINGS198955,Bali:2013fpo,Kalashnikova:2002zz,Grach:2008ij,Caselle:2013qpa,Johnson:2000qz,Kalaydzhyan:2014tfa, Caselle:818185}. Despite of the close relevance of to full QCD, detailed verification of the string conjecture is still incomplete in particular on distance scales before the string breaks~\cite{Bali:2005bg}. 
    




  Numerical examinations ~\cite{Giudice:2009di,Caselle:2004jq} do not seem to support the relevance of higher-order~\cite{Bicudo:2018yhk} model-dependent corrections for the NG action~\cite{Caselle:2004jq, Caselle:2004er} to the deviation from the free string model. These include 3D percolation model ~\cite{Giudice:2009di}, $Z(2)$, $SU(2)$ and $SU(3)$ confining gauge models~\cite{Caselle:2004jq}.
    
  Many features of the fine structure of the profile of QCD flux-tube at high temperature and relatively short distances ~\cite{Bakry:2018kpn, Bakry:2017utr,Brandt:2017yzw, Brandt:2010bw, Dubovsky:2013gi} are hoped to be compatible with modifications dictated by considering other effects beyond the free approximation. The delicate effects of emerging from the symmetry breaking near the edges or the string's resistance to bending, or in other words possible rigidity properties~\cite{POLYAKOV19, Kleinert:1986bk,Ambjorn:2014rwa}, suggest the boundary terms and/or the geometrical terms suppressing the sharp fluctuation in LW action, respectively. 

  The Lorentz-invariant boundary corrections~\cite{Caselle:2014eka,Aharony:2010cx, Aharony:2009gg, Aharony:2011gb} to the static $Q\bar{Q}$ potential~\cite{Caselle:2014eka,Brandt:2017yzw,Bakry:2017utr} have been addressed in simulations involving different operators and gauge groups~\cite{Caselle:2014eka, Brandt:2017yzw, Bakry:2017utr}. In particular, the fine deviations from the effective string and numerical outcomes~\cite{Billo:2011fd} could be explained following this reasoning.

  The broadening profile of the energy field ought to receive similar corrections from these Lorentzian-invariant boundary terms in the action. The contributions of the boundary action to the width profile is calculated recently in Ref.~\cite{Bakry:2019}. The modification to the mean-square width around the free NG string is evaluated using perturbative expansion of two boundary terms at the orders of fourth and six derivatives.

  Recent observation concerning the simulation of baryonic flux tubes~\cite{Bakry:2014gea,Bakry:2016uwt,Bakry:2011kga} indicates resistance to bending or repulsion among the flux lines at a junction of the network such that the angles between the three flux tubes are kept equally divided into $120^{o}$. This would suggest that the sharply-creased worldsheet configurations are energtically unfavorable. This interpretation of the self-repulsion or resistance to bending nature appears within the vortex line picture~\cite{PhysRevD.51.1842, Deldar:2009aw} of the confining string appears as well.

  The smooth string model rigorously preserves the fundamental properties of QCD of the ultraviolet (UV) freedom and infrared (IR) confinement~\cite{POLYAKOV19,Kleinert:1986bk}. The model is in consistency with glueballs~\cite{Kleinert:1988hz,Viswanathan:1987et} formation, and a real ($Q\bar{Q}$) potential~\cite{German:1989vk,Nesterenko1992,Ambjorn:2014rwa} with a possible tachyonic free spectrum~\cite{Kleinert:1996ry} above some critical coupling~\cite{Viswanathan:1987et}.
      
  The assumption of smooth QCD strings have consequences that could clearly manifest in the intermdiate distances at high temperatures or for the excited states. Though not dominating the IR region, rigidity effects are reported~\cite{Caselle:2014eka,Brandt:2017yzw,Bakry:2017utr} in recent numerical simulation of abelian and non-abelian gauge groups.
         
  Recently a set analytic solutions targeting the profile of flux-tube have appeared in the literature~\cite{allais,Gliozzi:2010zt,Bakry:2017fii,Bakry:2019,Giataganas:2015yaa}. The target of the present paper is to examine in detail the physical implications of each string model against the lattice numerical data.

   The map of the paper is as follows: In section(II), we review the most relevant string model to QCD corresponding to the energy width versus different approximation schemes. In section(III), lays out the numerical discussion of the lattice data with verious combinations of the effective strings models. In the last section, we provide concluding remarks. 
  
   \section{String actions and energy width}
  
   It is a reverberate conjectured~\cite{Nambu:1974zg,Nambu1979372} that the gluonic field may condenses into thin stringlike object that can admit a long string description. The string may follow from the intuition picture that the chromo-electric field is squeezed by the dual Miesener effect similar to Abriksov line in the superconductive scenario ~\cite{Mandelstam76, Bali1996, DiGiacomo:1999a, DiGiacomo:1999b, Carmona:2001ja, Caselle:2016mqu, Pisarski} of QCD vacuum. 

  The dynamical description of the effective string is based on a low-energy effective Lagrangian respecting the symmetries of the system. The classical long string solution, even though, breaks the translational invariance of the Yang-Mills vacuum leading to generation of transverse massless Goldston bosons~\cite{goddard, Low:2001bw}.
  
  The L\"uscher-Wiesz effective action includes all massless fields which are necessarily derivatives to impose the translational invariance and are expressed in the physical gauge~\cite{Dubovsky:2012sh,Aharony:2011ga}. The Lorentz-invariance of the LW action is realized nonlinearly sice the worldsheet gauge diffeomorphism is fixed to static/physical gauge.
  
  The L\"uscher and Weisz~\cite{Luscher:2004ib} (LW) effective action up to four-derivative term read
  
\begin{equation}
\begin{split}
  S_{LW}[\mathbf{X}]&=S_{cl}+\dfrac{\sigma}{2} \int d^{2} \zeta \Big(\dfrac{\partial \mathbf{X}}{\partial \zeta_{\alpha}} \cdot \dfrac{\partial \mathbf{X}}{\partial \zeta_{\alpha}}\Big)\\
 & +\sigma \int d^{2} \zeta  \Bigg[c_2 \Big( \dfrac{\partial \mathbf{X}}{\partial \zeta_{\alpha}} \cdot \dfrac{\partial \mathbf{X}}{\partial \zeta_{\alpha}} \Big)^2 + c_3 \Big( \dfrac{\partial \mathbf{X}}{\partial \zeta_{\alpha}} \cdot \dfrac{\partial \mathbf{X}}{\partial \zeta_{\beta}} \Big) ^2 \Bigg]\\
  &+\gamma \int d\zeta^2 \sqrt{g} \mathcal{R}+ \alpha\int d\zeta^2 \sqrt{g} \mathcal{K}^2+...+S_{b},
\label{LWaction}
\end{split}
\end{equation}
   in the above $S_{cl}$ is the classical term, the operators $X^{\mu}(\zeta^{0},\zeta^{1})$ define the mapping from $ C \subset \mathbb{R}^{2} $ into $\mathbb{R}^{4}$ taken with an Euclidean signature. The geometrical invariant $\mathcal{R}$ and $\mathcal{K}$ are the Ricii-scalar and the extrinsic curvature ~\cite{POLYAKOV19, Kleinert:1986bk} of the worldsheet configuration, respectively.
  The LW action Eq.~\eqref{LWaction} encompasses built-in surface/boundary terms to account for an open string with boundaries. The boundary action $S_{b}$  is located at the boundaries $\zeta^{1}=0$ and $\zeta^{1}=R$.



  The kinematic couplings $c_1$, $c_2$ are dependent and are subject to constraint
\begin{equation}
c_2 + c_3 = \dfrac{-1}{8\sigma}.
\label{couplings2}
\end{equation}

  which follows from the Lorentz-transform in terms of the string collective variables $X_i$ ~\cite{Aharony:2009gg,Billo:2012da} that the action is invariant under $SO(1,D-1)$.
  
   The Nambu-Goto action, however, is the most simple form of string actions and is proportional to area of the world-sheet. With the the above condition \eqref{couplings2}, the first two terms in LW action coincide with the leading and next to leading order terms of NG action given by
  
\begin{align}
 S^{\rm{NG}}_{\rm{\ell o}}[X]=\sigma_{0} A+\dfrac{\sigma_{0}}{2} \int d\zeta^{2} \left(\dfrac{\partial \bm{X} }{\partial \zeta_{\alpha} } \cdot \dfrac{\partial \bm{X} }{\partial \zeta_{\alpha}} \right), 
\label{NGLO}
\end{align}     
and
\begin{align}
S^{\rm{NG}}_{\rm{n\ell o}}[X]= \sigma_{0} \int d\zeta^{2} \left[ \left(\dfrac{\partial \bm{X}}{\partial \zeta_{\alpha}} \cdot \dfrac{\partial \bm{X}}{\partial \zeta_{\alpha}}\right)^2 + \left(\dfrac{\partial \bm{X}}{\partial \zeta_{\alpha}} \cdot \dfrac{\partial \bm{X}}{\partial \zeta_{\beta}}\right)^2\right],
\label{NGNLO}
\end{align}
  respectively.

  The quantum delocalization of the stringlike flux-tube around its classical configuration results in an energy distribution profile along the line connecting two color charges. The second moment of the transverse fluctuations typically characterizes the mean-square width of the string
  
\begin{align}
W^{2}(\zeta;\zeta_{0}) = & \quad \langle \, X^{2}(\zeta;\zeta_{0})\,\rangle \nonumber\\ 
                = &\quad \dfrac{\int_{\mathcal{C}}\,[D\,X]\, X^2 \,\mathrm{exp}(-S_{\rm{eff}}[X])}{\int_{\mathcal{C}}[D\,X] \, \mathrm{exp}(-S_{\rm{eff}}[X])},
\label{StringWidth} 
 \end{align}
\noindent where $\zeta=(\zeta_{1},i\zeta_{0})$ is a complex parameterization of the cylindrical worldsheet of surface area $R\,L$, $S_{eff}$ denote general effective string action.

  The Dirichlet and periodic boundary condition in $\zeta_{0}$ with period $L_{T}$ corresponds to
  
\begin{equation}
\begin{split}
 X(\zeta_{1},\zeta_{2}=0)&= X(\zeta_{1},\zeta_{2}=R)=0,\\ 
 X (\zeta_{1}=0,\zeta_{2})&= X (\zeta_{1}=L_T=\frac{1}{T},\zeta_{2}).
\end{split}
\label{DPBC}
\end{equation}

  The leading order perturbative solution of Eq.~\eqref{StringWidth} subject to the boundary condition Eq.~\eqref{DPBC} revealed the famed logarithmic divergence of the width at the middle of the string and at zero temperature which is the famed property shown by L\"uscher, M\"unster and Weisz~\cite{Luscher:1980iy} long ago.

\begin{equation}
\label{Wid}  
  W^{2}_{NG_{\ell o}} \sim \frac{1}{\pi\sigma_{0}}\log(\dfrac{R}{R_{0}}),
\end{equation}
\noindent where $R_{0}$ is an ultraviolet (UV) scale. 

  Allais and Casselle ~\cite{allais, Gliozzi:2010zv} using point-split~\cite{Caselle:1995fh} regularization and conformal mapping techniques have evaluated the expectation value of the quardratic operator Eq~\eqref{StringWidth} at any temperature and plane in accord to  
\begin{equation}
\label{WidthLO}
W^{2}_{{\ell o}}(\zeta,\tau) = \frac{D-2}{2\pi\sigma_{0}}\log\left(\frac{R}{R_{0}(\zeta)}\right)+\frac{D-2}{2\pi\sigma_{0}}\log\left| \,\dfrac{\theta_{2}(\pi\,\zeta/R;\tau)} {\theta_{1}^{\prime}(0;\tau)} \right|,
\end{equation}
 which corresponds to the Green function of the free bosonic string theory in two dimensions. The $\theta$ functions in Eq.~\eqref{WidthLO} are Jacobi elliptic functions defined as
\begin{eqnarray}
\theta_{1}(\zeta;\tau)=2 \sum_{n=0}^{\infty}&(-1)^{n}q_1^{n(n+1)+\frac{1}{4}}\sin((2n+1)\,\zeta),\nonumber\\
\theta_{2}(\zeta;\tau)=2 \sum_{n=0}^{\infty}&q_1^{n(n+1)+\frac{1}{4}}\cos((2n+1)\zeta),
\end{eqnarray}
\noindent , $q_1=e^{\frac{-\pi}{2}\tau}$, $\tau=\frac{L_T}{R}$ is the modular parameter of the cylinder, and $L_T=1/T$ is the temporal extent governing the inverse temperature and $R_{0}(\zeta)$ is the UV cutoff which has been generalized to be dependent on distances from the sources. 

  At high temperature the long string limit $R>L_T$ of Eq.~\eqref{WidthLO} implies a linear broadening pattern in the string's width~\cite{allais,Gliozzi:2010zv}. The second logarithmic term in Eq.~\eqref{WidthLO} implies a different width at each plane around the middle of the string. This curved form becomes more pronounced with the increase of the temperature and the string's length.

  F. Gliozzi. M. Pepe and Wiese \cite{Gliozzi:2010zv,Pepe:2010na} extended the calculations of the width to two-loop order of perturbative expansion of NG action Eq.~\eqref{NGNLO}, the next-to-leading  width reads as
    
\begin{equation}
W^2(\zeta)=W^2_{\ell o}(\zeta)+W^{2}_{n\ell o}(\zeta)
\end{equation}
with the leading order term $W^2_{\ell o}$ in accord to Eq.~\ref{WidthLO} and the NLO term given by
\begin{equation}
\begin{split}
\label{WidthNLO}
W^{2}_{n\ell o}(\zeta)=&\frac{\pi}{12 \sigma_{0} R^2} \left[E_2(i\tau)-4E_2(2i\tau)\right]\left( W_{lo}^2(\xi) -\frac{D-2}{4\pi \sigma_{0}}\right)\\
&+\dfrac{(D-2)\pi}{12\sigma_{0}^2 R^2} \Big\{\tau \left(q \frac{d}{dq}-\frac{D-2}{12}E_2(i\tau)\right) \\
&\times \left[E_2(2i\tau)-E_2(i\tau)\right]-\frac{D-2}{8 \pi} E_2(i\tau)\Big\},
\end{split}
\end{equation}

  where $q=e^{-\pi \frac{L}{R} }$. The form of $W^2_{\ell o}$ in terms of Dedekind $\eta$ function given in Ref.~\cite{Gliozzi:2010zv} is equivalent to Eq.~\eqref{WidthLO} through the standard relations of elliptic functions.

where Eisenstein series $E_{2}$ defined as 
\begin{equation}
  E_{2}(\tau) = 1 -24 \sum_{n=1}^{\infty} \frac{n q^{n}}{1-q^n}.
\label{E2}
\end{equation}
and Dedkind $eta$ function is defined as
\begin{equation}
  \eta(\tau)=q^{\frac{1}{24}} \prod_{n=1}^{\infty}(1-q^{n}).
\label{eta}  
\end{equation}

  As mentioned above, an interesting generalization of the Nambu-Goto string ~\cite{Arvis:1983fp, Alvarez:1981kc, Olesen:1985pv} has been proposed by Polyakov~\cite{POLYAKOV19} and Kleinert~\cite{kleinert} to stabilize the NG action in the context of fluid membranes. The Polyakov-Kleinert string is a free bosonic string with additional Poincare-invariant term proportional to the extrinsic curvature of the surface as a next order operator after NG action~\cite{kleinert, POLYAKOV19}. That is, the surface representation of the Polyakov-Kleinert (PK) string depends on the geometrical configuration of the embedded sheet in the space-time. That is, the bosonic free string action is equiped with additional terms of the extrinsic curvature as a next-order operator after NG action~\cite{POLYAKOV19, Kleinert:1986bk}.

   Many properties have been rigorously worked out by Kleinert and German such as the exact potential in the large dimension limit~\cite{Braaten:1987gq,Kleinert:1989re}, the dynamical generation of the string tension~\cite{Kleinert:1988vq} and the perturbative stability in critical dimensions~\cite{PhysRevLett.58.1300}. This is in addition to various thermodynamical characteristics of the geometric strings gas including the partition function~\cite{Elizalde:1993af}, free energy and string tension at finite temperature~\cite{Viswanathan:1988ad,German:1991tc,Nesterenko:1997ku} and the deconfinement transition point~\cite{Kleinert:1987kv}.

   The smooth configurations of quantum fluctuations swept in the Euclidean space-time by the Nambu-Goto string are favored in the string's partion function by adding a new term proportional to the geometrical second fundamental form, or simply extrinsic curvature of the worldsheet.

   The second fundamental form (or the shape tensor) in a differential geometer notation defines a quadratic form on the tangent plane of a smooth surface in the three-dimensional Euclidean space. With a smooth choice of the unit normal vector at each point, this quadratic form is generalized as a smooth hypersurface in a Riemannian manifold.

   The action of the Polyakov-Kleinert (PK) string with the extrinsic-curvature term reads
\begin{equation}
  S^{\rm{PK}}[X]= S_{\rm{\ell o}}^{\rm{NG}}[X]+S^{R}[X],
\label{PKaction}    
\end{equation}
  with $S^{R}$ defined as 
\begin{equation}
  S^{\rm{R}}[X]=\alpha_{r} \int d^2\zeta \sqrt{g}\, {\cal K}^2.
\label{Ext}  
\end{equation}  
  The extrinsic curvature ${\cal K}$ defined as 
\begin{equation}
{\cal K}=\triangle(g) \partial_\alpha [\sqrt{g} g^{\alpha\beta}\partial_\beta],
\end{equation}
  where $\triangle$ is Laplace operator and $M^{2}=\frac{\sigma_{0}}{2 \alpha_{r}}$ is the rigidity parameter. The term satisfies the Poincare and the parity symmetries and can also be considered~\cite{Caselle:2014eka} in the general class of (LW) actions~\eqref{LWaction}.

  The perturbative expansion ~\cite{German:1989vk} of the rigidity term Eq.~\eqref{Ext}
  
\begin{equation}
S^{\rm{R}}[X]=S_{\rm{\ell o}}^{R}[X]+S_{\rm{n \ell o}}^{R}[X]+...,    
\end{equation}

has the leading term given by

\begin{equation}
S_{\rm{\ell o}}^{R}= \alpha_{r} \int_{0}^{L_T} d\zeta_0 \int_{0}^{R} d\zeta_1  \left[\left(\dfrac{\partial^{2} \bm{X}} {\partial \zeta_{1}} \right)^2 + \left(\dfrac{\partial^{2} \bm{X}} {\partial \zeta_{0}^{2}}  \right)^2 \right]
\label{S_extlo}
\end{equation}

and the next to leading-order term

\begin{equation}
\begin{split}
  S_{\rm{n\ell o}}^{R}=& \int_{0}^{L_T} d\zeta_0 \int_{0}^{R} d\zeta_1 \bigg[\dfrac{\sigma}{8}\left( \dfrac{\partial X} {\partial \zeta_{\alpha}} \right)^4 -\dfrac{\sigma}{4} \left( \dfrac{\partial \bm{X}} {\partial \zeta_{\alpha}} \cdot \dfrac{\partial \bm{X}} {\partial \zeta_{\beta}}  \right)^2\\
    &+ 2 \alpha_{r} \dfrac{\partial^{2} \bm{X}} {\partial \zeta_{\alpha} \partial \zeta_{\beta}} X^{2}- \dfrac{\alpha_{r}}{2} \left(\dfrac{\partial \bm{X}}{\partial \zeta_{\alpha}} \cdot \dfrac{\partial^{2} \bm{X}} {\partial \zeta_{\beta}^{2}} \right)^2 \\
&-\alpha_{r} \left( \dfrac{\partial \bm{X}}{\partial \zeta_{\alpha}}\cdot \dfrac{\partial \bm{X}}{\partial \zeta_{\beta}} \right) \left( \dfrac{\partial^{2} \bm{X}} {\partial \zeta_{\alpha} \partial \zeta_{\beta}} \cdot \dfrac{\partial^{2} \bm{X}}{\partial \zeta_{\gamma}^{2}} \right)\bigg]. 
\label{S_extnlo}
\end{split}
\end{equation}

   The rigidity parameter weighs favorably the smooth worldsheet configuration over the creased one. In non-abelian gauge theories this ratio is expected to remain constant in the continuum limit~\cite{Caselle:2014eka}.

   The numerical simulations~\cite{Caselle:2014eka} of the confining potential in $U(1)$ gauge theory have first addressed the rigidity of the effective bosonic string. Possible manifestation in $SU(N)$ gauge theories in $3D$~\cite{Brandt:2017yzw} has been reported as well. The rigidity effects in the confining potential of $SU(3)$ at high temperature manifests as a necessary ingradient to retrieve the correct dependency of the string tension on the temperature~\cite{Bakry:2017utr,Bakry:2018kpn}.

   
   The mean-squared width of the Polyakov-Kleinert string can be calculated by expanding around the free-string action Eq.~\eqref{NGLO} the squared width of the string
   
\begin{equation}
W^2(\zeta) = W_{\ell o}^2(\zeta)+ \langle  X(\zeta, \zeta_{0})^2  S^{R} \rangle_0 +...
\label{TwoLoopExpansion}
\end{equation}
  where $< >_0$ represents the vacuum expectation value with respect to the free-string action. The modification to the mean-square width by virtue of the leading term in the rigidity

\begin{equation}
W^2_{R_{(\ell o)}}= \left\langle  X^2(\zeta,\zeta_{0})^2 S^{R}_{\ell o} \right\rangle,
\label{Width_Rig_LO}  
\end{equation}
  In the following, we evaluate the correlator Eq.~\eqref{Width_Rig_LO} of the rigid string up to one loop order using Green function 
\begin{equation}
G(\zeta,\zeta_{0};\zeta',\zeta_{0}^{'})=\left \langle X(\zeta,\zeta_{0}) X(\zeta',\zeta_{0}^{'}) \right \rangle.
\end{equation}  
as the two point propagator. On a cylindrical sheet of surface area $RL$ with  Dirichlet and periodic boundary condition in $\zeta_{0}$ with period $L_{T}$ the Green propagator of the free string is
\begin{equation}
  \begin{split}
G(\zeta,\zeta_{0};\zeta',\zeta^{\prime}_{0})&=  \frac{1}{\pi \sigma_{0} }\sum _{n=1}^{\infty } \frac{\sin \left(\frac{\pi  n \zeta}{R}\right) \sin \left(\frac{\pi  n \zeta'}{R}\right)}{n \left(1-q^n\right)}\\
&\times \left(q^n e^{\frac{\pi  n (\zeta_{0}-\zeta_{0 \prime})}{R}}+e^{-\frac{\pi n (\zeta_{0}-\zeta_{0 \prime})}{R}}\right),
\end{split}
\label{FP}
\end{equation}

  Equation~\eqref{Width_Rig_LO}, representing the perturbation in the width due to rigidity around the free NG string, in terms of the corresponding Green functions is

\begin{equation}
\begin{split}
  &\left\langle X(\zeta,\zeta_{0})^2 S^{R} \right\rangle=\\
  &(D-2)\lim_{\epsilon, \epsilon' \to 0 }\int d^2\zeta'(\partial_{\mu}^{2} G(\zeta;\zeta^{\prime}) \partial_{\mu'}^{2}G(\zeta;\zeta^{\prime})).\\
\label{Integral}
\end{split}
\end{equation}

  The modification to the mean-square width of the string of smoothed fluctuation is calculated in detail in Ref.~\cite{Bakry2017fii} using $\zeta$ function regularization technique and turn out to be

\begin{equation}
\label{WExt}  
W_{\rm{R_{(\ell o)}}}^2=\frac{-\pi(D-2)\alpha_{r}}{24 R^2 \sigma ^2 }  E_2 \left( \frac{ L_{T}}{2 R}\right),
\end{equation}

  The mean-square width of the rigid string at the next to leading term in the perturbative expansion of extrinsic curvature Eq.~\eqref{S_extnlo} is explicitly calculated in Ref.~\cite{Bakry2017fii}, the two-loop version of Eq.~\eqref{WExt} read as

\begin{equation}
\begin{split}
  W^{2}_{\rm{R_{(n\ell o)}}}&= \frac{ \pi^3  \alpha_r (3(D-2)^2-(D-2))}{8 R^5 \sigma^3}\\
  &\Bigg( \frac{4 R}{\pi} \left(\log \left( \eta \left(\tau/2  \right)\right)+\gamma \right)+\frac{L_{T}}{12} \left(2-E_{2}\left(\tau/2 \right) \right) \Bigg)\\
  &\left(\frac{1}{240}E_4\left(\tau\right)+\frac{1}{240}\right)+\frac{\pi^2 \alpha_{r}(D-2)}{8 R^4 \sigma^3}.
\end{split}
\label{Ext2loopT}
\end{equation}

  Apart from the possible stiff structure of QCD strings, the symmetry breaking of the action at the boundaries can have detectable effects on the energy density along the QCD flux tube as well. This perturbation from the free bosonic string behavior has been discussed in the numerical data of static potential ~\cite{Billo:2012da, Brandt:2017yzw, Bakry:2017utr,talk}.


A Generic boundary action can be defined as

\begin{equation}
S_{b}=\int d\zeta^0 \left[\mathcal{L}_{1}+\mathcal{L}_2+\mathcal{L}_3+\mathcal{L}_4+...   \right],
\label{GBaction}
\end{equation}

  with the Lagrangian density $\mathcal{L}_i$ associated with the corresponding effective low-energy parameter $b_{i}$. Dirichlet boundary conditions  $X^i=0$ at both ends means that $\zeta^0$-derivatives vanish on the boundary ($\partial_0^n X=0$). At the lowest order, the only possible term \cite{Luscher:2002qv} in the Lagrangian is therefore
 

\begin{align}
\label{L1}
\mathcal{L}_1 = b_1 {\partial_1\mathbf{X}} \cdot {\partial_1\mathbf{X}}.
\end{align}

The leading-order corrections due to second boundary terms with the coupling $b_2$ appears at the four-derivative term in the bulk. On the boundary the general term is of the form $\partial^{3} {\mathbf{X}}^2$

\begin{align}
\label{L2}
\mathcal{L}_2=b_2 {\partial_0 \partial_1 \mathbf{X}} \cdot {\partial_0\partial_1 \mathbf{X}}~,
\end{align}

one should note that possible terms proportional to the equation of motion can be set to zero by field redefintion.

The third term of coupling $b_3$ is given by
\begin{equation}
\mathcal{L}_3= b_3 \left(\partial_1 \mathbf{X} \cdot \partial_1 \mathbf{X} \right)^2.
\label{L3}
\end{equation}

 The Lorentz symmetry is a crucial aspect of Yang-Mills theory that ought to be preserved. The application of the Lorentz-transformation on the boundary action Eq.~\eqref{GBaction} and requiring $S_{b_i}$ to vanish, we obtain constraints~\cite{Aharony:2011gb,Gliozzi:2011hj} on the values of the couplings; or realize higher-order derivatives in the choice of the action of a given coupling.


  The variation of the boundary actions at first and third orders with infinitesimal nonlinear Lorentz-transform ~\cite{Billo:2012da} of the Lagrangian densities Eq.\eqref{GBaction} entails vanishing value for $b_1=0$ and $b_3=0$. It can be shown~\cite{Luscher:2004ib} that $b_1=0$ based on duality in two different channels corresponding open-closed string.
    
  The nonlinear realization of Lorentz transformation of the Lagrangian densities Eq.~\eqref{L2} and Eq.~\eqref{L4} generates higher-order terms at the same scaling~\cite{Aharony:2011gb, Billo:2012da}. The invariance of \eqref{L2} leads to recursion relation when solved Ref.\cite{Billo:2012da} give rise to a more general form of the Lagrangian density which encompasses the naive constructs Eq.~\eqref{L2} as special cases,

\begin{equation}
     \mathcal{L}_2=b_2  \left( \dfrac{\partial_0 \partial_1 \mathbf{X} \cdot \partial_0 \partial_1 \mathbf{X}}{1+\partial_1 \mathbf{X} \cdot \partial_1 \mathbf{X}}-\dfrac{(\partial_0 \partial_1 \mathbf{X} \cdot \partial_1 \mathbf{X})^2}{(1+\partial_1 \mathbf{X} \cdot \partial_1 \mathbf{X})^2} \right).
     \label{b2Complete}
\end{equation}

  The next Lagrangian density $\mathcal{L}_4$ of coupling $b_4$, the leading general effective Lagrangian on the boundary is

\begin{equation}
\mathcal{L}_4=b_4 \partial_{0}^2\partial_{1} \mathbf{X} \cdot \partial_{0}^2\partial_{1} \mathbf{X}.
\label{L4}
\end{equation}

 The first two-terms derived in Ref.\cite{Billo:2012da} are given by

\begin{equation}
\begin{split}   
   \mathcal{L}_4&=b_4 \Bigg( \frac{\partial_{0}^2\partial_{1} \mathbf{X} \cdot \partial_{0}^2\partial_{1} \mathbf{X} }{1+\partial_1 \mathbf{X} \cdot \partial_1 \mathbf{X}}-\\
   &\dfrac{(\partial_0^{2}\partial_1 \mathbf{X} \cdot \partial_1 \mathbf{X})^2+4 (\partial_{0}^{2}\partial_{1}\mathbf{X} \cdot \partial_{0}\partial_{1} \mathbf{X})(\partial_{0}\partial_{1} \mathbf{X})}{(1+\partial_1 \mathbf{X} \cdot \partial_1 \mathbf{X})^2}+...\Bigg).
\label{b4Complete}
\end{split}
\end{equation}

  The boundaries do affect the average width of the delocalization along the string. The mean-square width is similarly expressed as perturbation around the free NG string as

\begin{equation}
W^2(\zeta) = W_{\ell o}^2(\zeta)+ \langle  X(\zeta, \zeta_{0})^2 (S_{b_2} + S_{b_4}+....) \rangle_0 +...,
\label{TwoLoopExpansion}
\end{equation}

The evaluation of leading correlator 

  \begin{equation}
  W^{2}_{b_2}=-\langle X^2 S_{b_2} \rangle.
\label{W2b2}  
  \end{equation}
  
  involve cumbersome manipulations, we presented the detailed calculus in Ref.\cite{Bakry:2019} using $\zeta$-function regularization of the divergent sums appearing after the evaluation of the Green propagator Eq.~\eqref{FP}
  
\begin{equation}
  W^{2}_{b_2}=\frac{-\pi b_2 (D-2)}{4 R^3 \sigma^2} \left(\frac{1}{8}-\frac{1}{24}E_{2}(\tau) \right), 
\label{1wb2}  
\end{equation}

where $E_2(\tau)$ is Eisenstein series defined by Eq.~\eqref{E2}. The expectation value of the six-derivative order boundary-term  

\begin{equation}
  W^{2}_{b_4}=-\langle X^2 S_{b_4}\rangle_{0},
\label{W2b4}  
\end{equation}

 is similarly evaluated substituting the free propagator Eq.\eqref{FP} (see Ref.~\cite{Bakry:2019} for detail). The next non-vanishing expectation value appears at the coupling $b_4$,

\begin{figure}[!hpt]
\begin{center}
\includegraphics[scale=0.35]{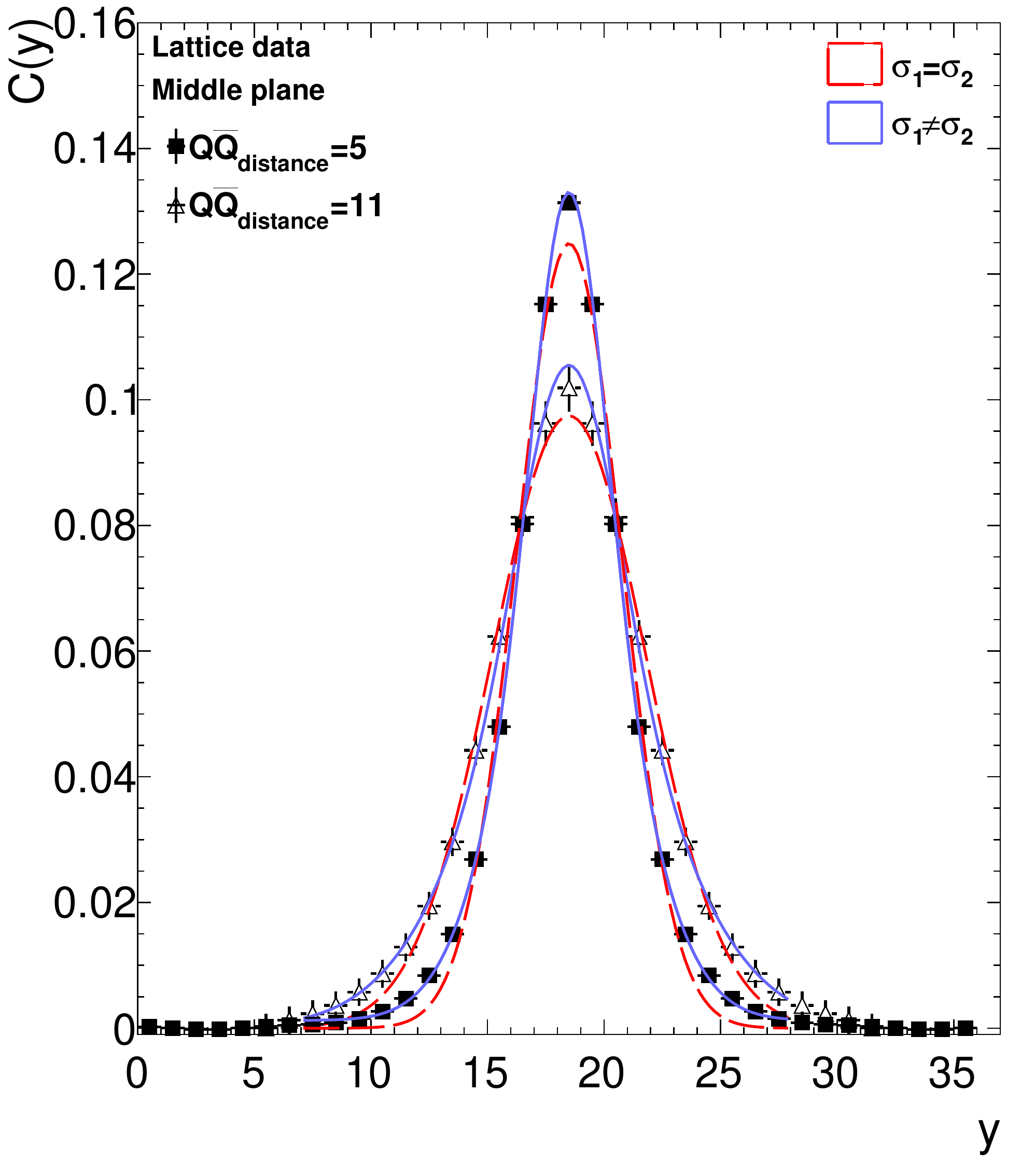} 
\caption{The density distribution $ \mathcal{C}(r,\theta,z=R/2)$ at the center of the tube $z=R/2$ for source separation  $R\,=0.5$ fm and $R\,=1.1$ fm at temperature $T/T_{c} \approx 0.9$. The solid and dashed lines correspond to the fit to Eq.\eqref{conGE}  with $\sigma_1\neq\sigma_2$ and $\sigma_1=\sigma_2$, respectively.
	}\label{action}
\end{center}
\end{figure}
\begin{figure*}[!hpt]
\begin{center}
\subfigure{\includegraphics[scale=0.26]{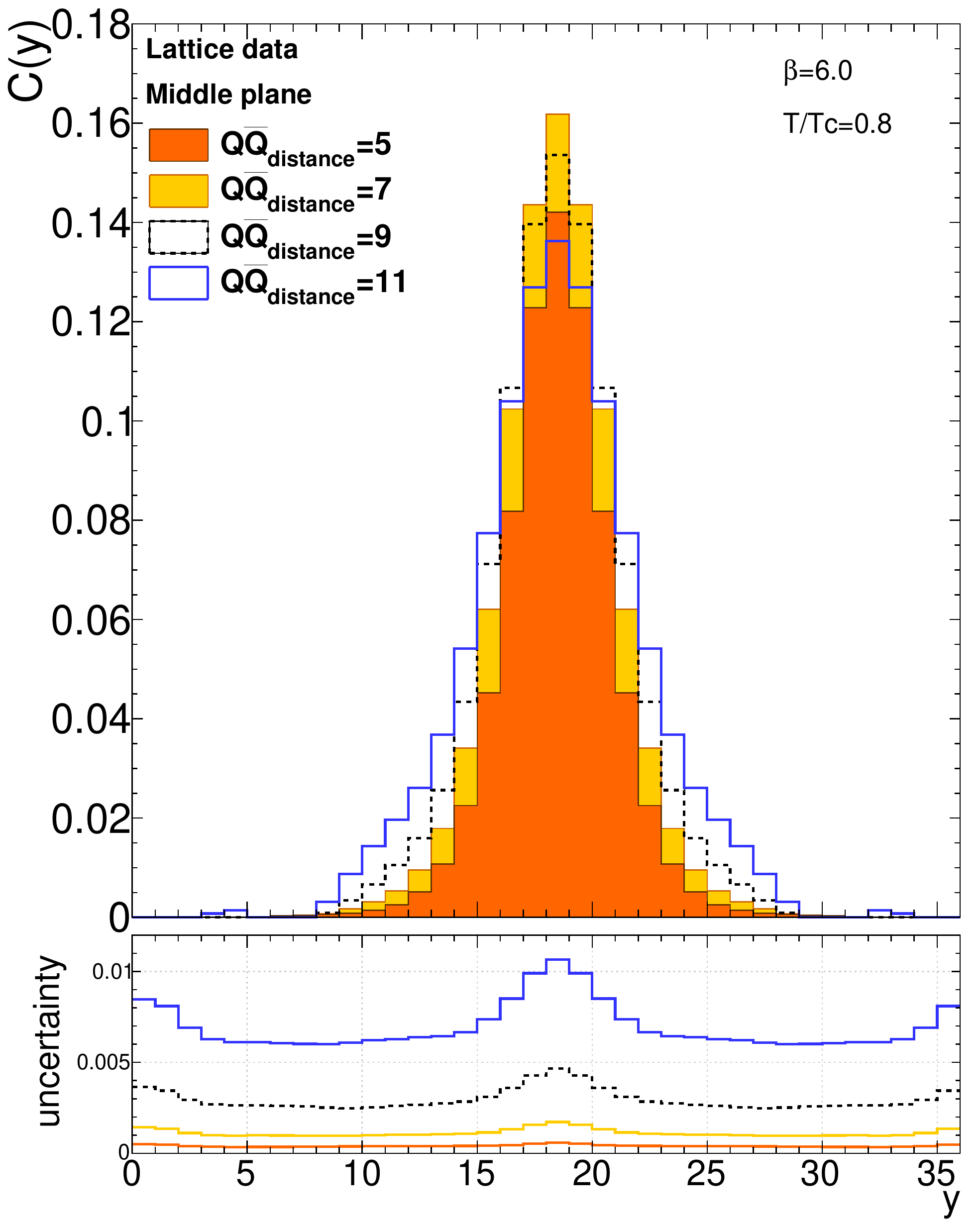}}
\subfigure{\includegraphics[scale=0.26]{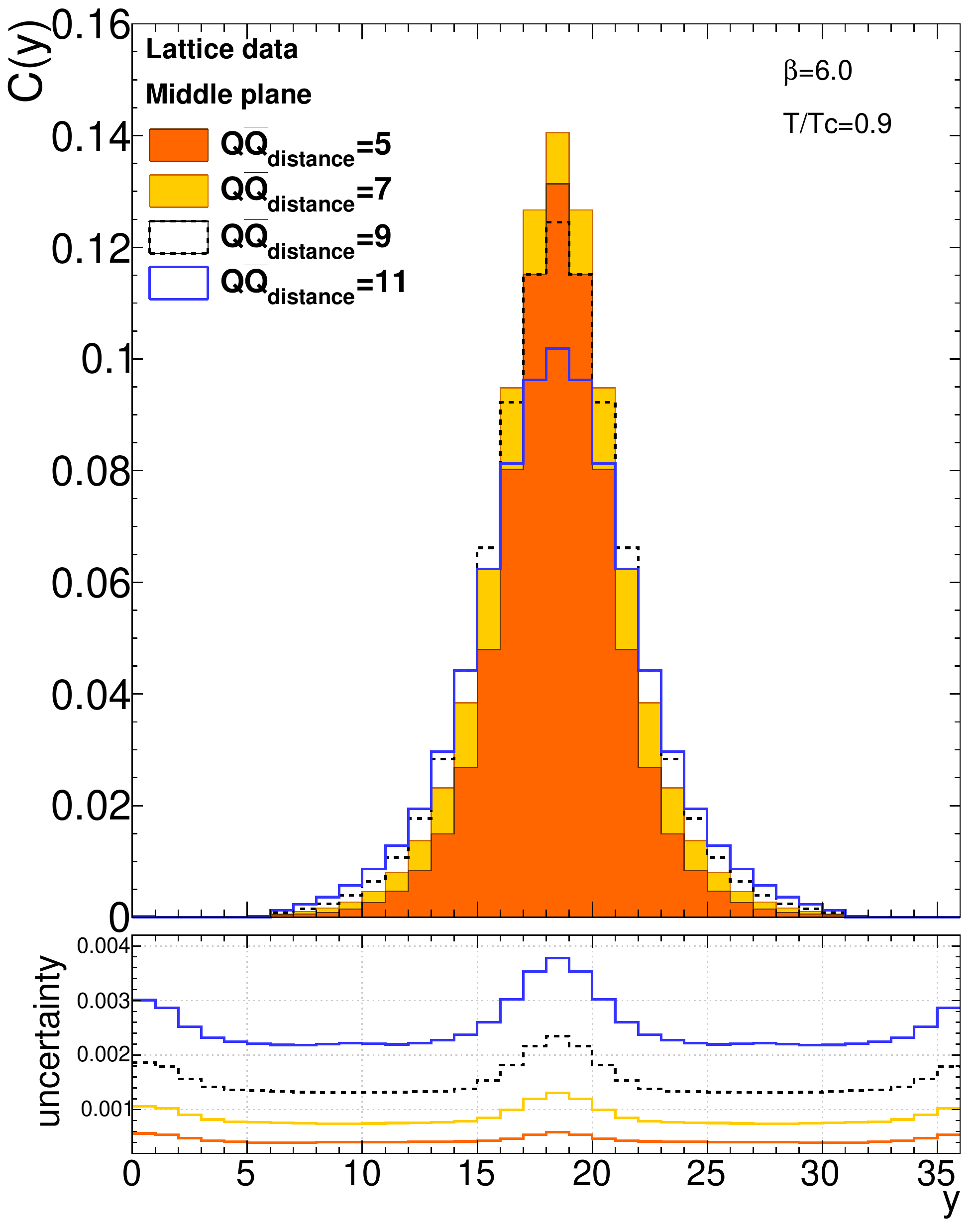}}
\subfigure{\includegraphics[scale=0.26]{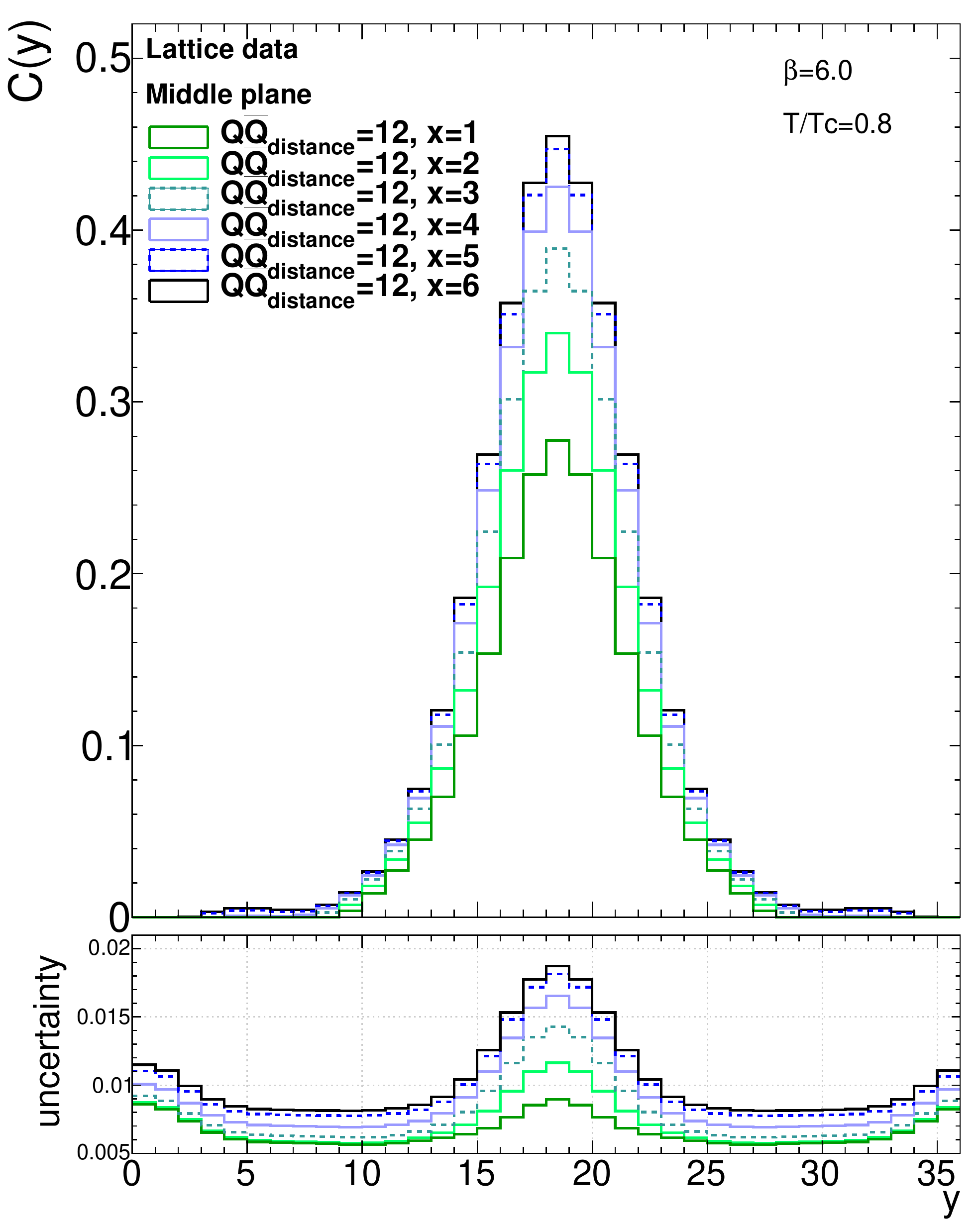}}
\caption{ The action density profile of quark-antiquark $Q\bar{Q}$ separation distances $R=5,7,9,11$ at the center of the tube, $z=R/2$. The pad below show uncertainty distributions of the corresponding action densities. Profile are shown for the depicted temperatures $T/T_{c}=0.8$ and  $T/T_{c}=0.9$.}\label{HIST0809}
\end{center}
\end{figure*}
\begin{figure*}[!hptb]
\begin{center}
\includegraphics[scale=0.62]{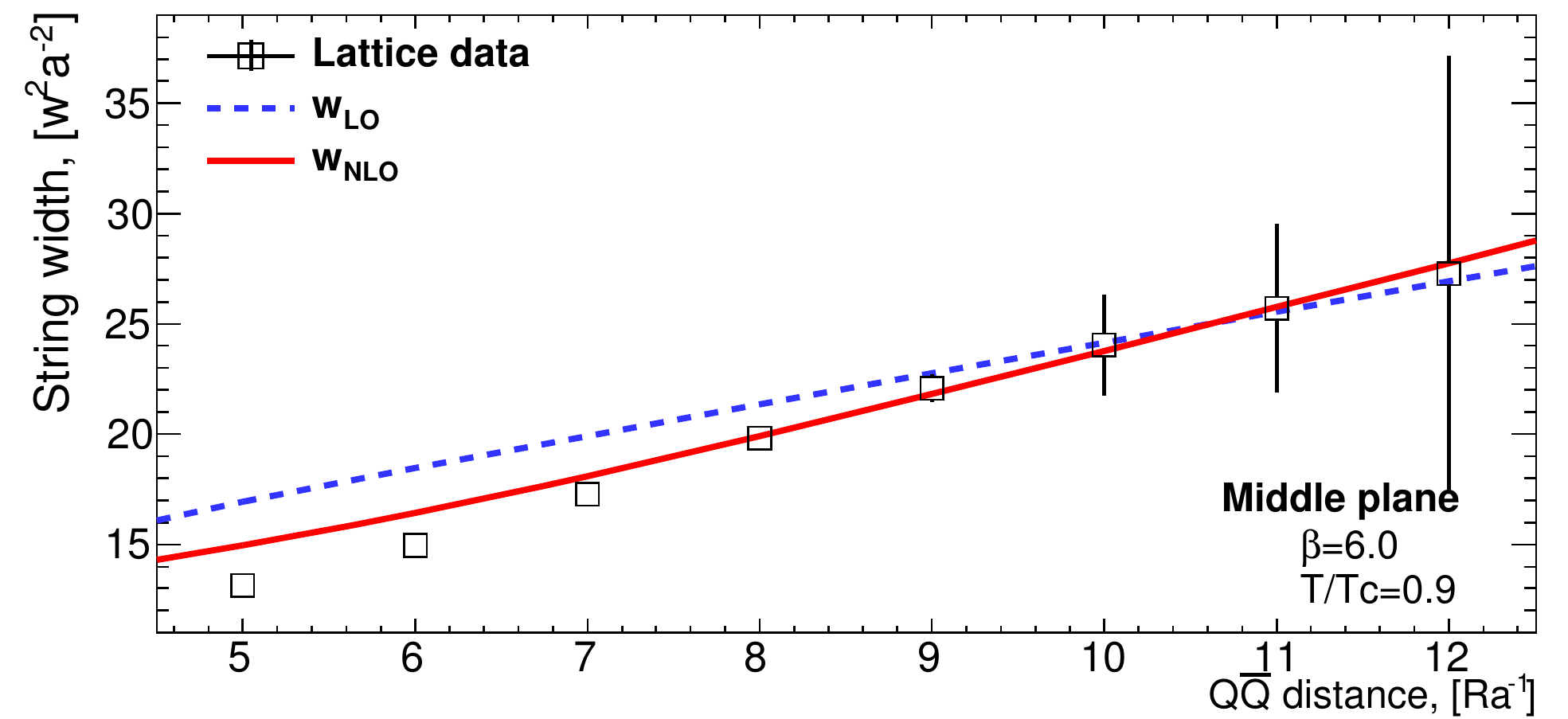}	
\caption{The mean-square width of the density distribution in the middle of the tube $z=R/2$ at the temperature $T/T_c \approx 0.9$. The solid and dashed lines correspond to the free and self-interacting NG string Eq.\eqref{WidthLO} and Eq.\eqref{WidthNLO}, respectively.}\label{MT09}
\end{center}
\end{figure*}
\begin{figure}[!hpt]
\begin{center}
\includegraphics[scale=0.56]{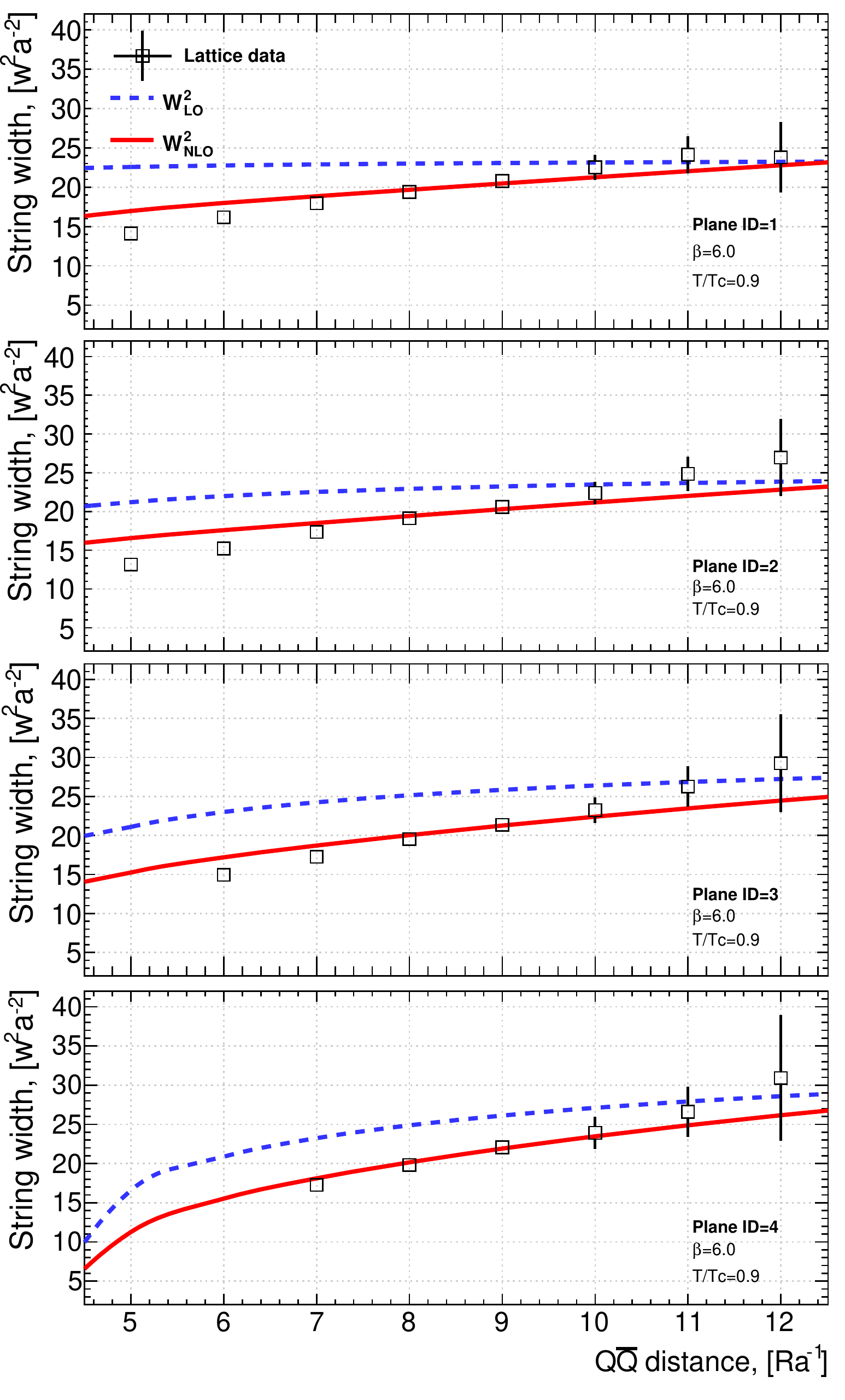}	
\caption{ The mean-square width $ W^{2}(z) $ of the string versus $Q\overline{Q}$ separations $R$ at temperature ~$ T/T_{c} \approx 0.9$ in lattice unit. Measurements are taken at consecutive planes $z=1$, $z=2$,~(c)~$z=3$ and ~$z=4$ from the the top to the bottom. The solid and dashed lines correspond to the one parameter fit to the string model, Eq.\eqref{WidthLO} and \eqref{WidthNLO}, respectively}\label{PlanesID1234T09}
\end{center}
\end{figure}
\begin{figure*}[!hpt]
\begin{center}
\includegraphics[scale=0.6]{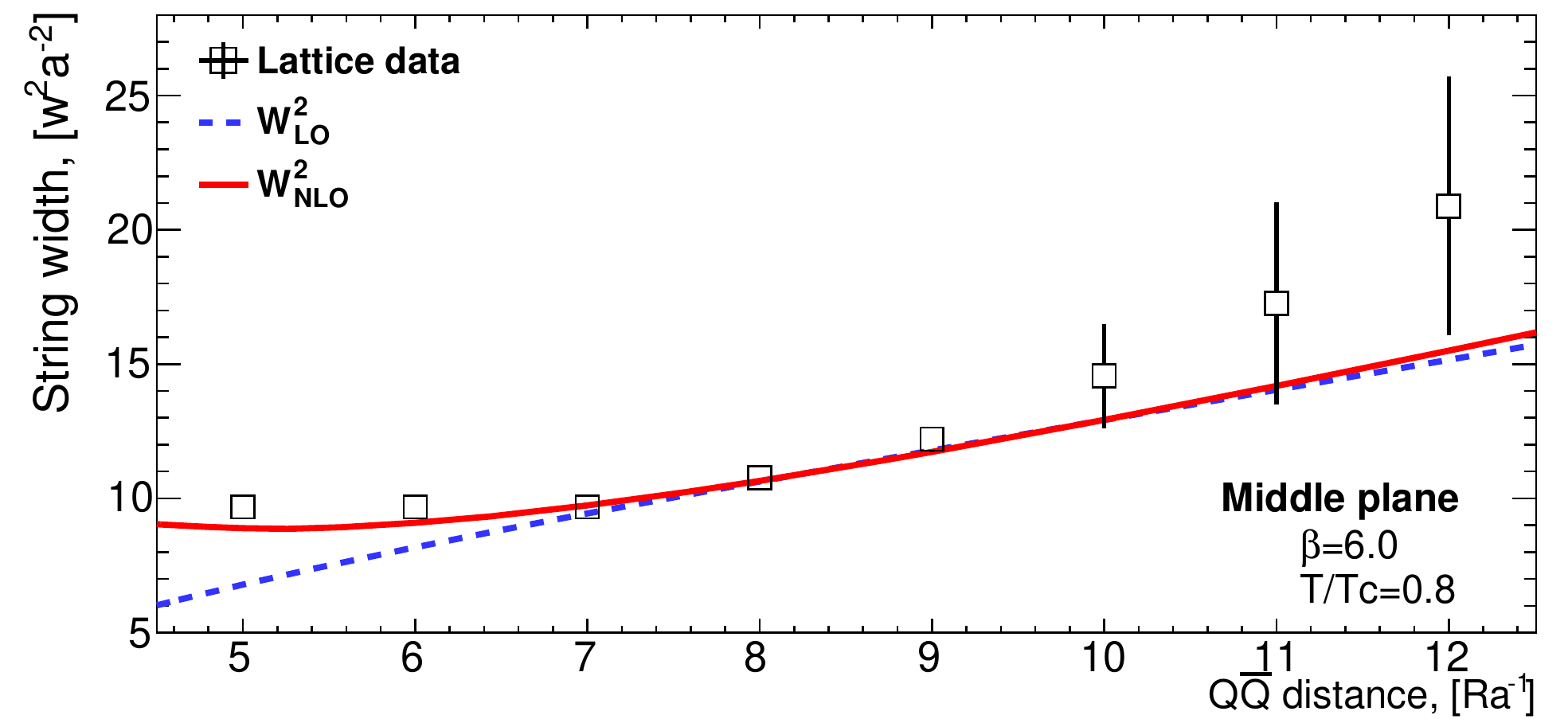}
\caption{The broadening of the mean-square width of the density-distribution at the center of the tube $z = R/2$ versus the $Q\bar{Q}$ separation distance $R$ at temperature $T/T_c=0.8$. The solid and dashed lines correspond to the free and self-interacting Nambu-Goto (NG) string (LO) Eq.~\eqref{WidthLO} and (NLO) Eq.~\eqref{WidthNLO}, respectively.}\label{MT08}
\end{center}
\end{figure*}
\begin{figure}[!hpt]
\begin{center}
\subfigure{
\includegraphics[scale=0.55]{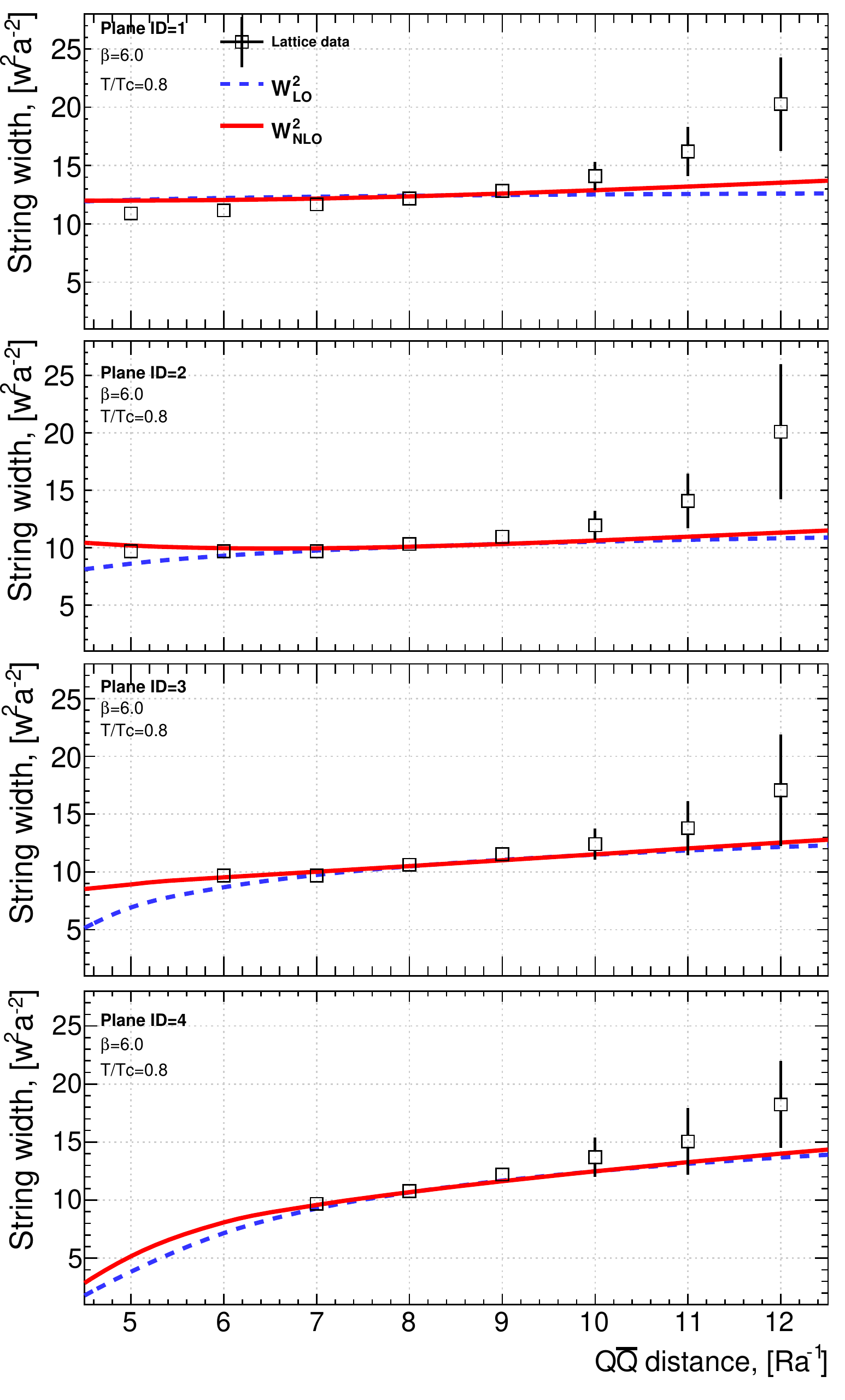}	
}
\caption{The mean-square width of the string $W^{2}(z)$ at ~$ T/T_{c} \approx 0.8$ versus $Q\overline{Q}$ separations measured in planes $z=1$,$z=2$, $z=3$, and $z=4$ from the top to bottom. The dashed and solid line denote the leading and next to leading order Nambu-Goto string model Eq.\eqref{WidthLO} and Eq.\eqref{WidthNLO}, respectively.}\label{PlanesID1234}
\end{center}
\end{figure}
\begin{figure}[!hpt]
\begin{center}
\subfigure[The width differences from the middle plane of tube at $R=0.8$ fm]{
\includegraphics[scale=0.4]{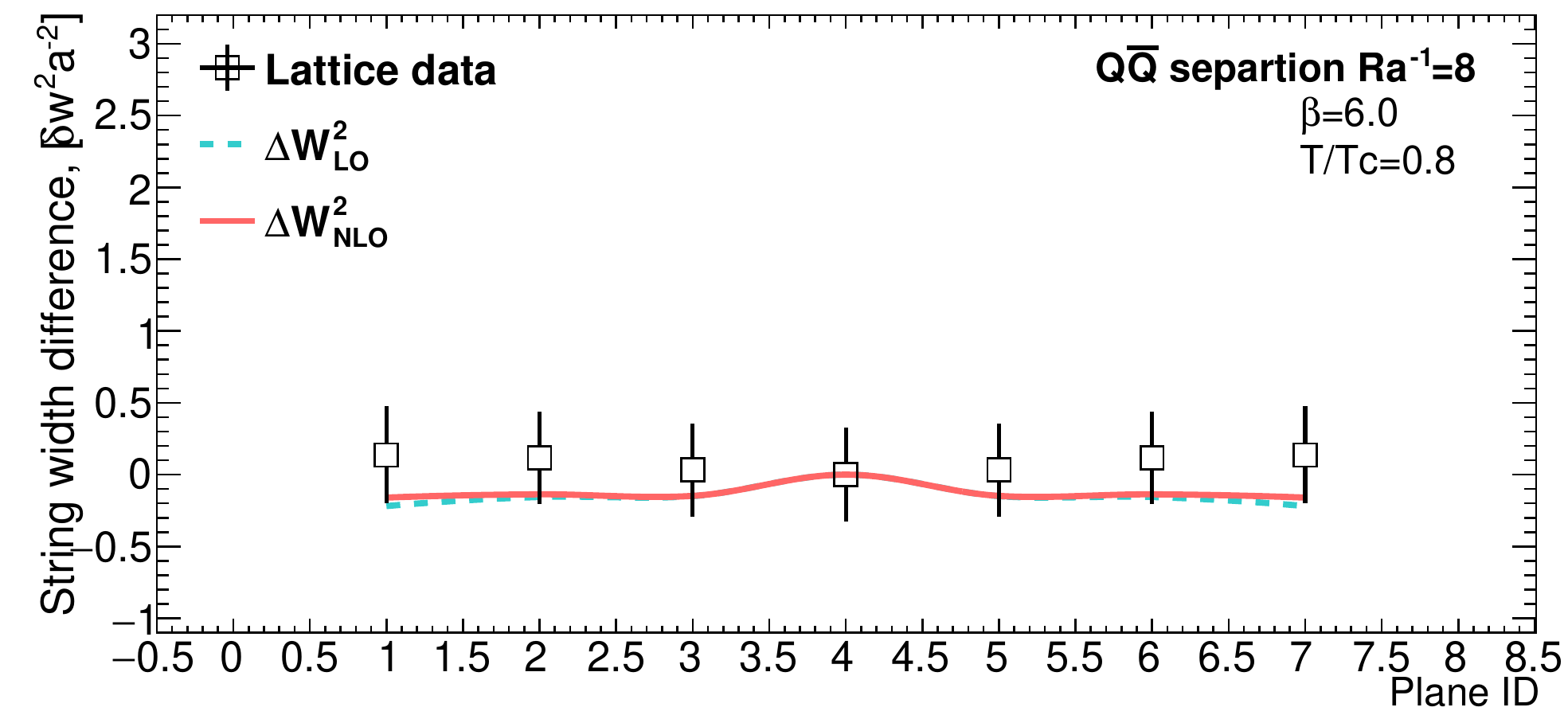}} 
\subfigure[Same as subfigure(a) at $R=0.9$ fm]{\includegraphics[scale=0.4]{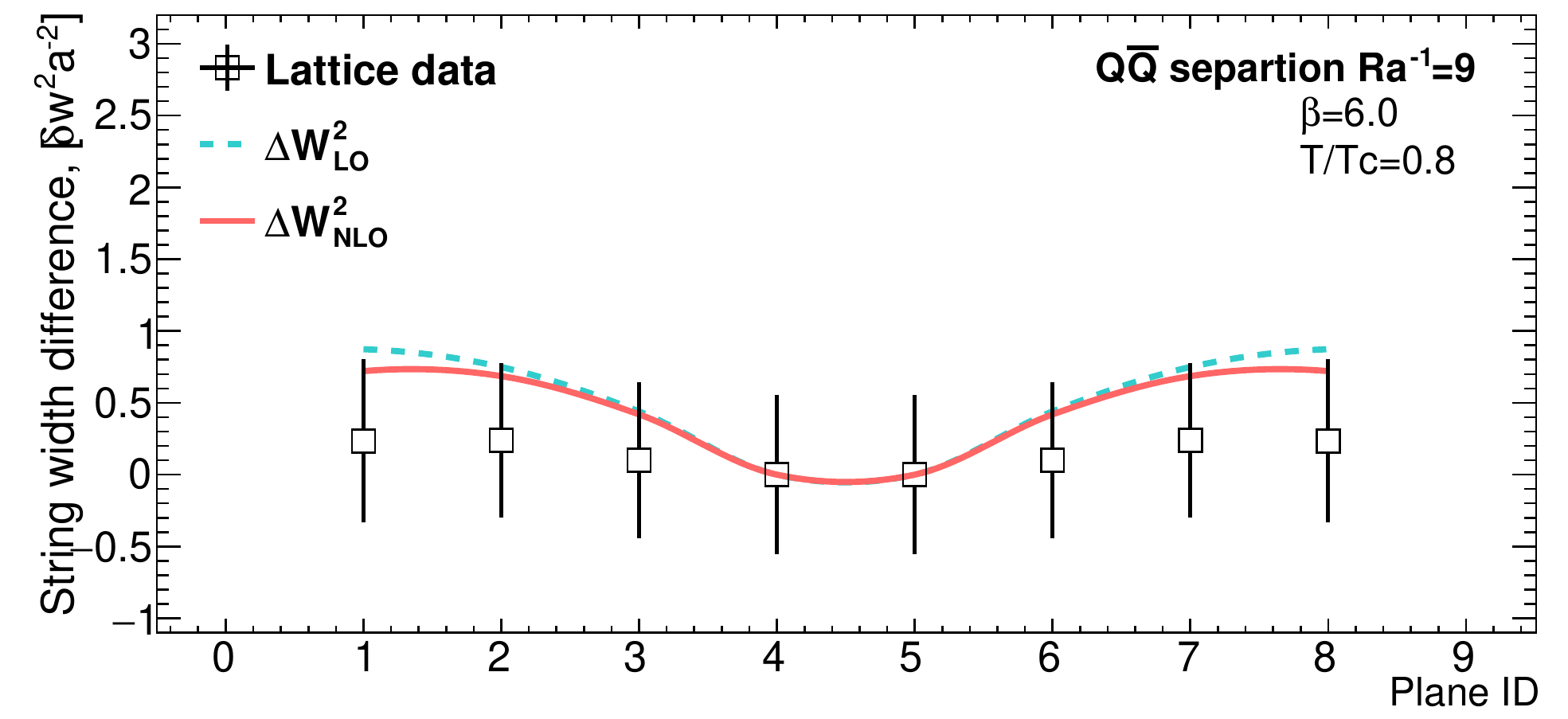}}  
\subfigure[The action density in the quark plane at $R=1.2$ fm]{\includegraphics[scale=0.39]{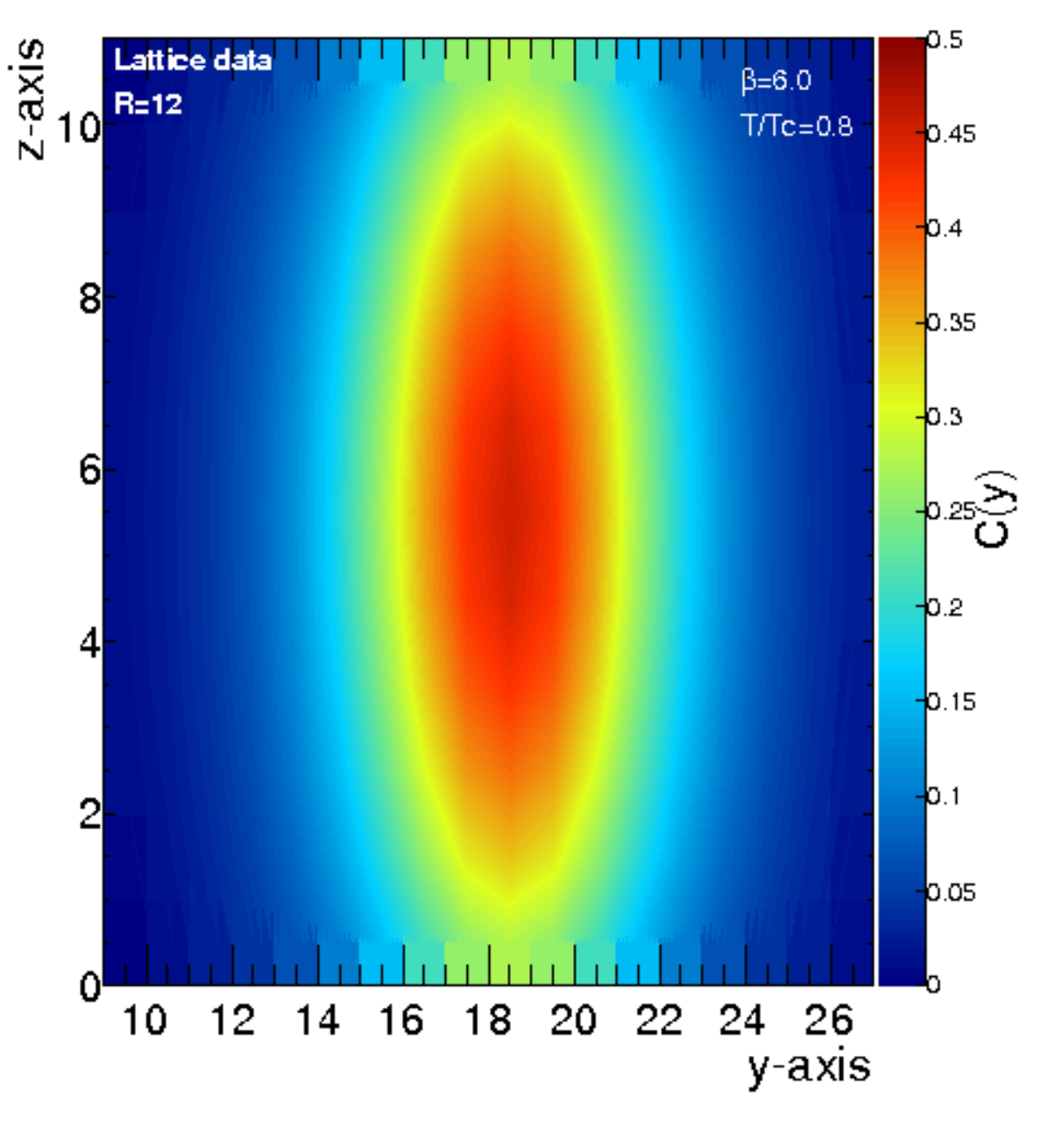}}
\subfigure[Same as subfigure (c) at $R=1.4$ fm]{\includegraphics[height=5.8cm, width=5.0cm]{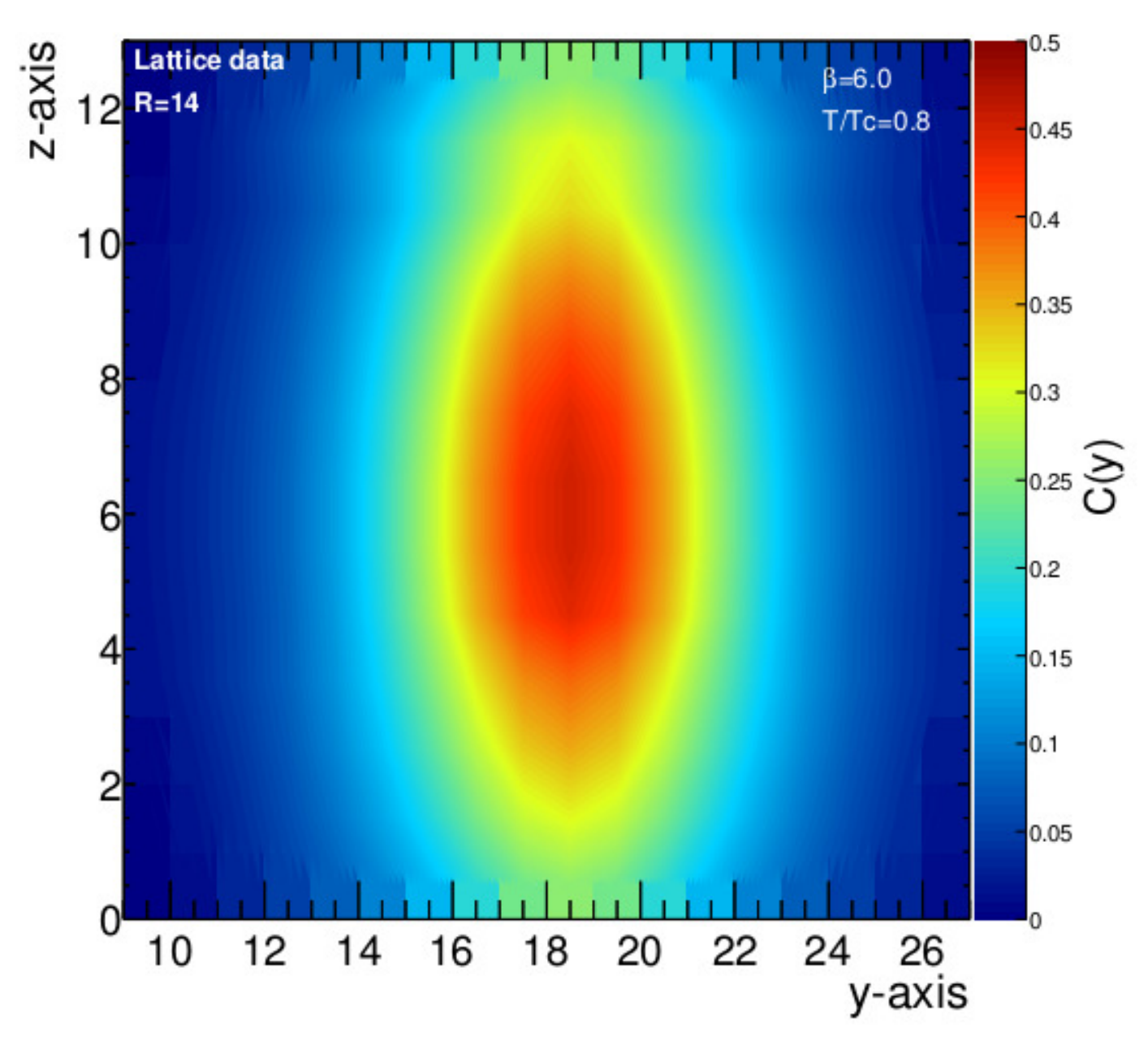}}
\caption{The density distribution is exhibiting a nonuniform pattern along the transverse planes, even though the tube's width is constant at all source separations.}\label{DIFFT08} 
\end{center}
\end{figure}
\begin{figure*}[!hpt]	
                        \includegraphics[scale=0.65]{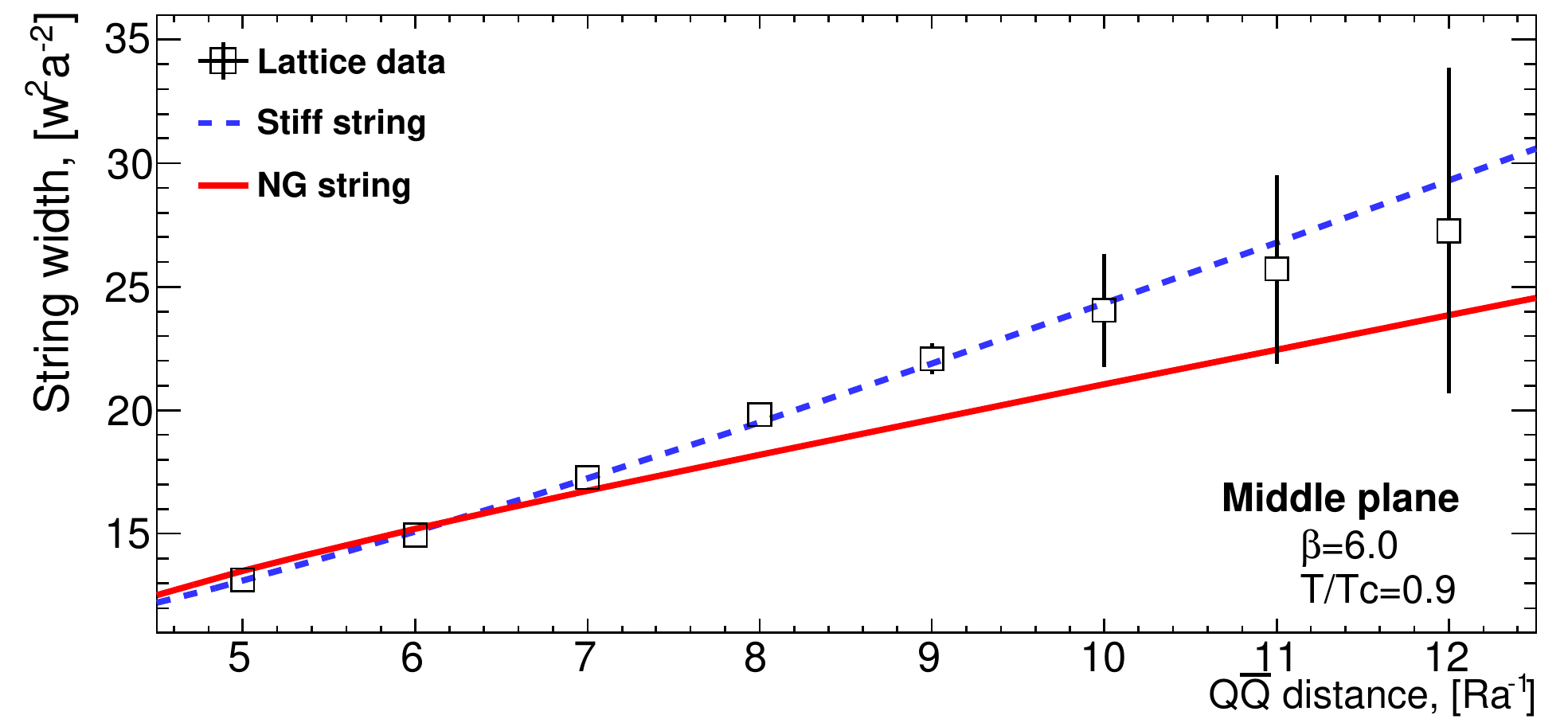}				
		\caption{
			The mean-square width of the string $ W^{2}(z) $ versus $Q\overline{Q}$ separations measured in the middle plane $R/2$ at ~$ T/T_{c} \approx 0.9$. The solid and dashed line denote are fits to Nambu-Goto and Stiff string Eqs.~\eqref{WidthNLO} and ~\eqref{WExt} on the interval $R\in[0.5,1.2]$ fm, respectively.}\label{MidPlane}		
\end{figure*}
 
\begin{figure}[!hpt]
			\includegraphics[scale=0.5]{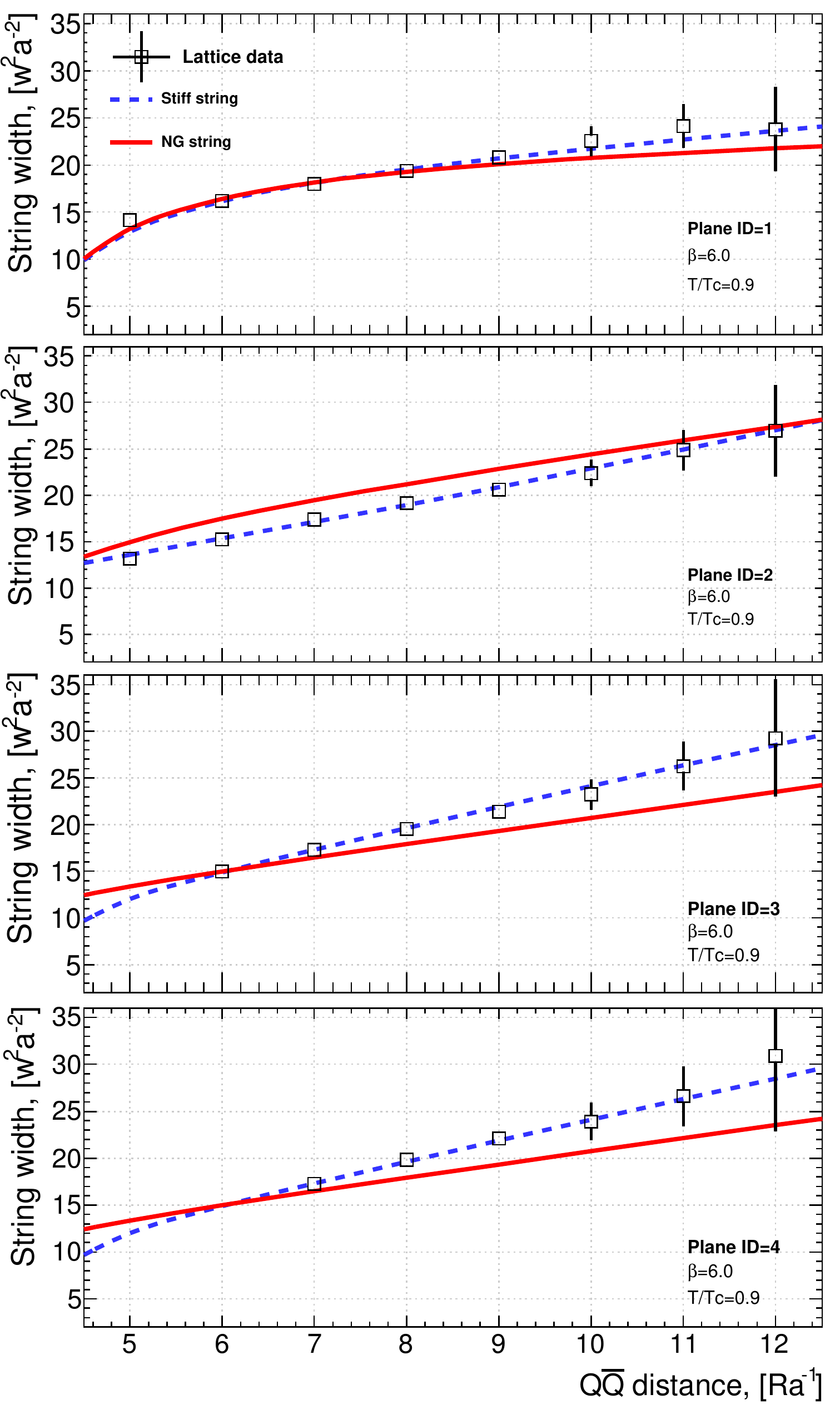}

		\caption{
			The mean-square width of the string $ W^{2}(z) $ versus $Q\overline{Q}$ separations measured in the planes $z=1$,$z=2$, $z=3$, and $z=4$ respectively from the top to bottom at  ~$ T/T_{c} \approx 0.9$. The solid and dashed line are the fits to Nambu-Goto and Rigid string Eqs.\eqref{WidthNLO} and \eqref{WExt} on the interval $R\in[0.5,1.2]$ fm, respectively.}\label{FitsPlanesR}		
\end{figure}

\begin{figure}[!hpt]	
\begin{center}						
\subfigure[]{\includegraphics[scale=0.4,angle=270]{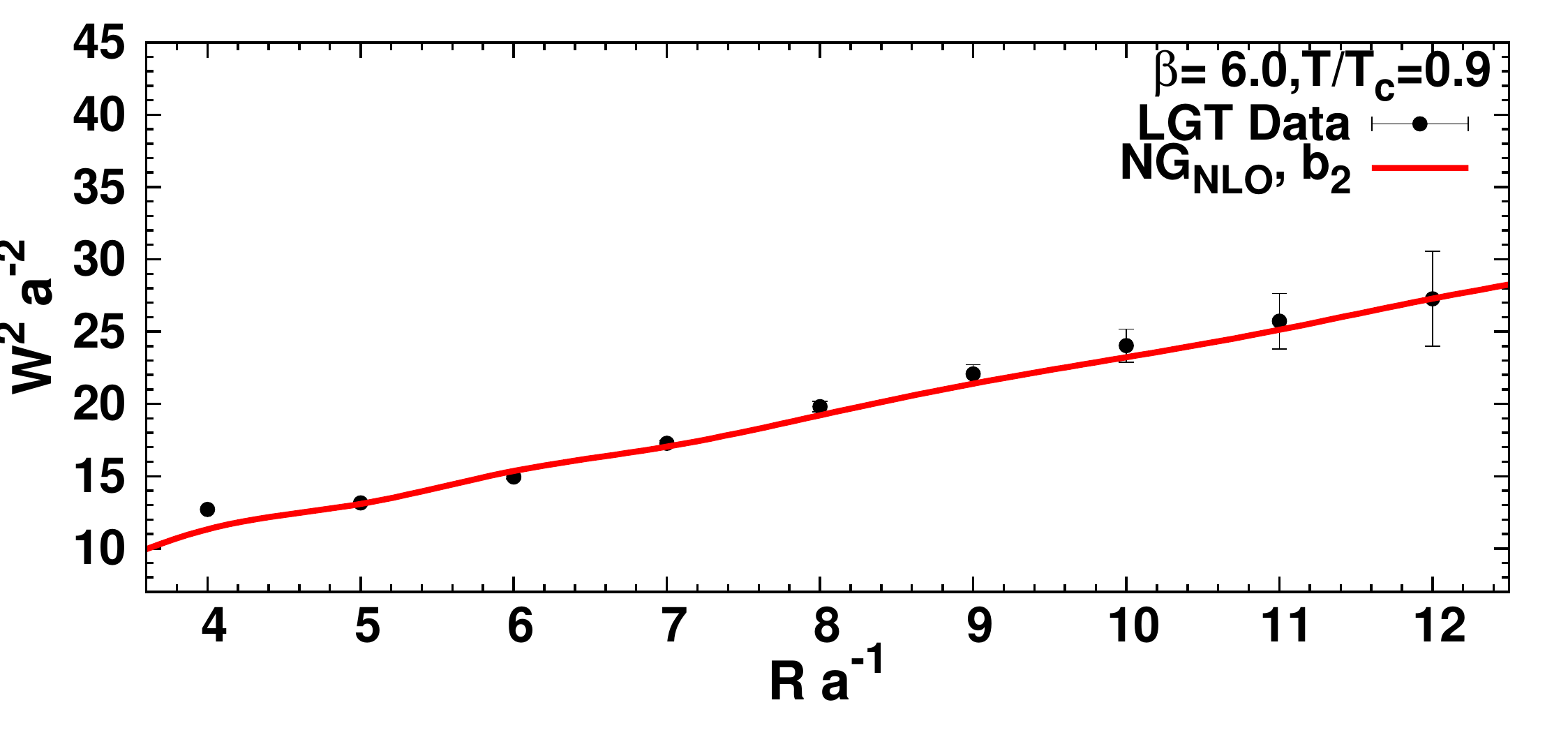}}
\subfigure[]{\includegraphics[scale=0.4,angle=270]{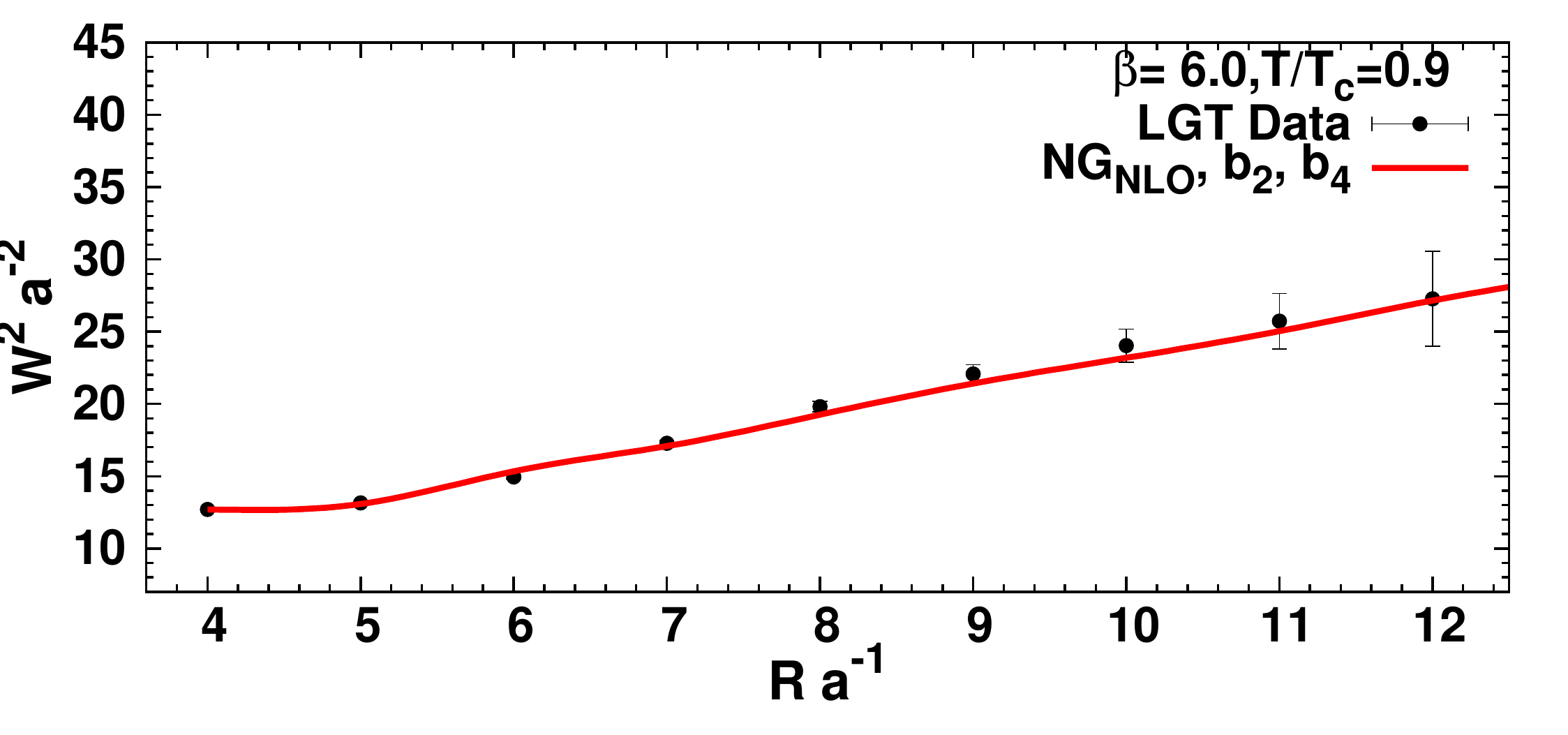}}
\caption{The mean-square width at the middle plane $R/2$ and temperature $T/T_{c} =0.9$. (a)Compares  the NG string at NLO with boundary terms at coupling $b_2$ Eq.~\eqref{Wid_NLO_Boundb2} and fit range $R \in [0.5,12] $ fm. (b)The same effective NG string; however, with two-boundary terms at coupling $(b_2, b4)$ Eq.~\eqref{Wid_NLO_Boundb2b4} for fit range $ R \in[0.4,1.2]$ fm.} \label{BoundFitMid}		
\end{center}						
\end{figure} 

\begin{figure}[!hpt]
 \subfigure[]{\includegraphics[scale=0.4,angle=270]{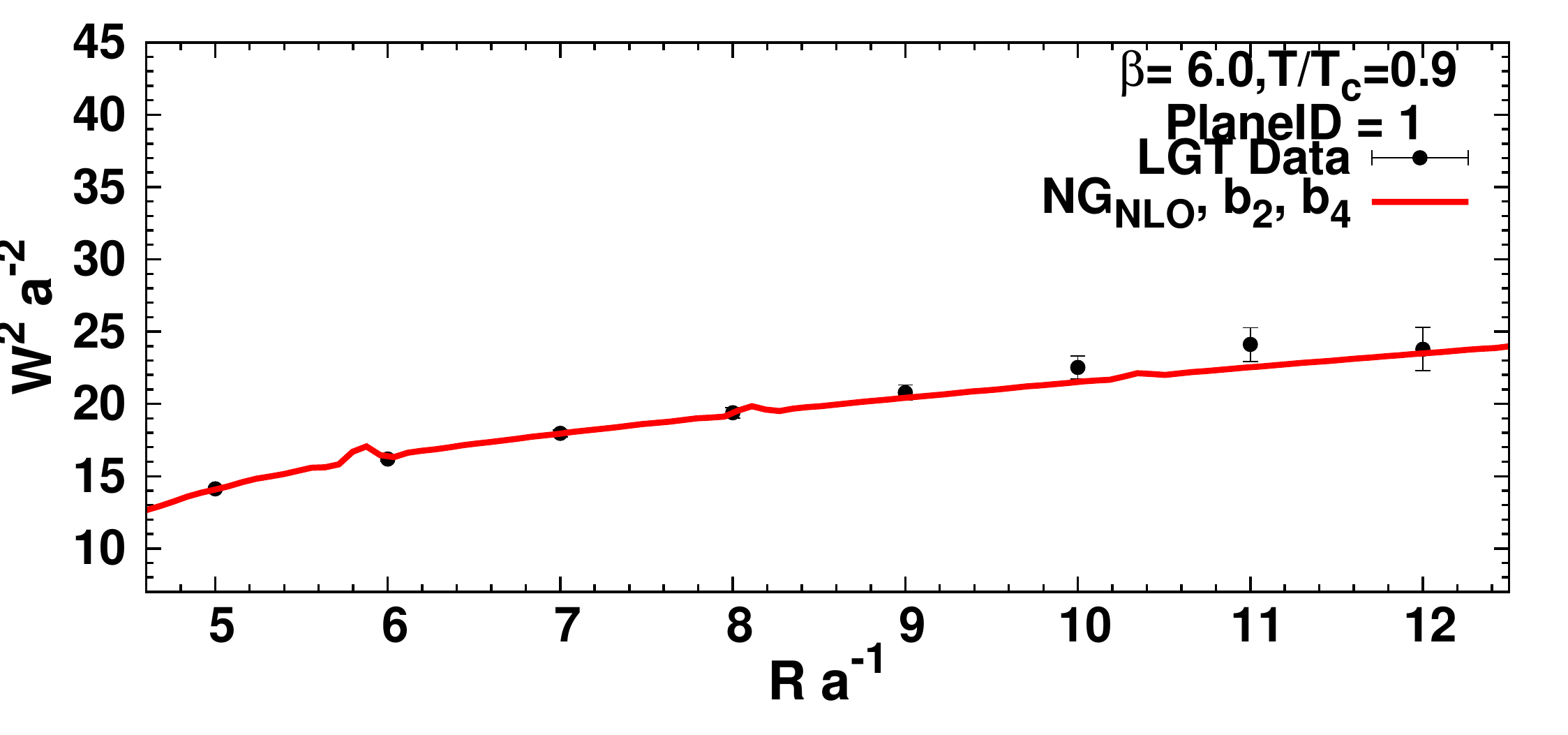}}
\subfigure[]{\includegraphics[scale=0.4,angle=270]{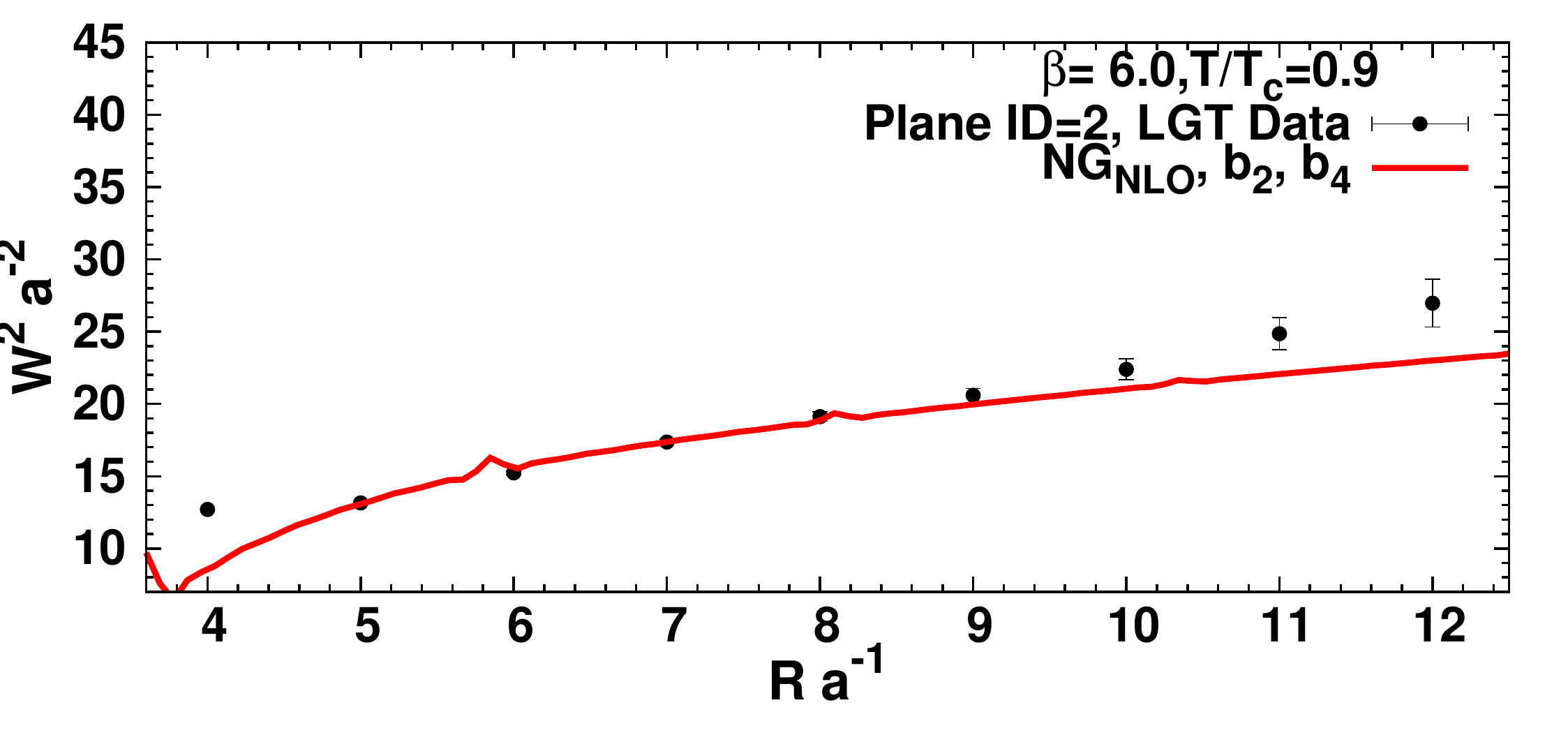}}
  \caption{The mean-square width of the string $W^2(z)$ versus $Q\bar{Q}$ separations measured at $T/T_c=0.9$. The lines correspond to the fits of NLO width of NG and string and the two-boundary of couplings $(b_2,b_4)$ Eq.~\eqref{Wid_NLO_Boundb2b4} on the interval $R\in[0.5,1.2]$ fm. (a) Width at the plane $z=1$(b) Width at the plane $z=2$.}\label{BoundFitPlanes}		
\end{figure}

\begin{figure}[!hpt]
\begin{center}
\subfigure[$R=0.8$ fm]{\includegraphics[scale=0.4]{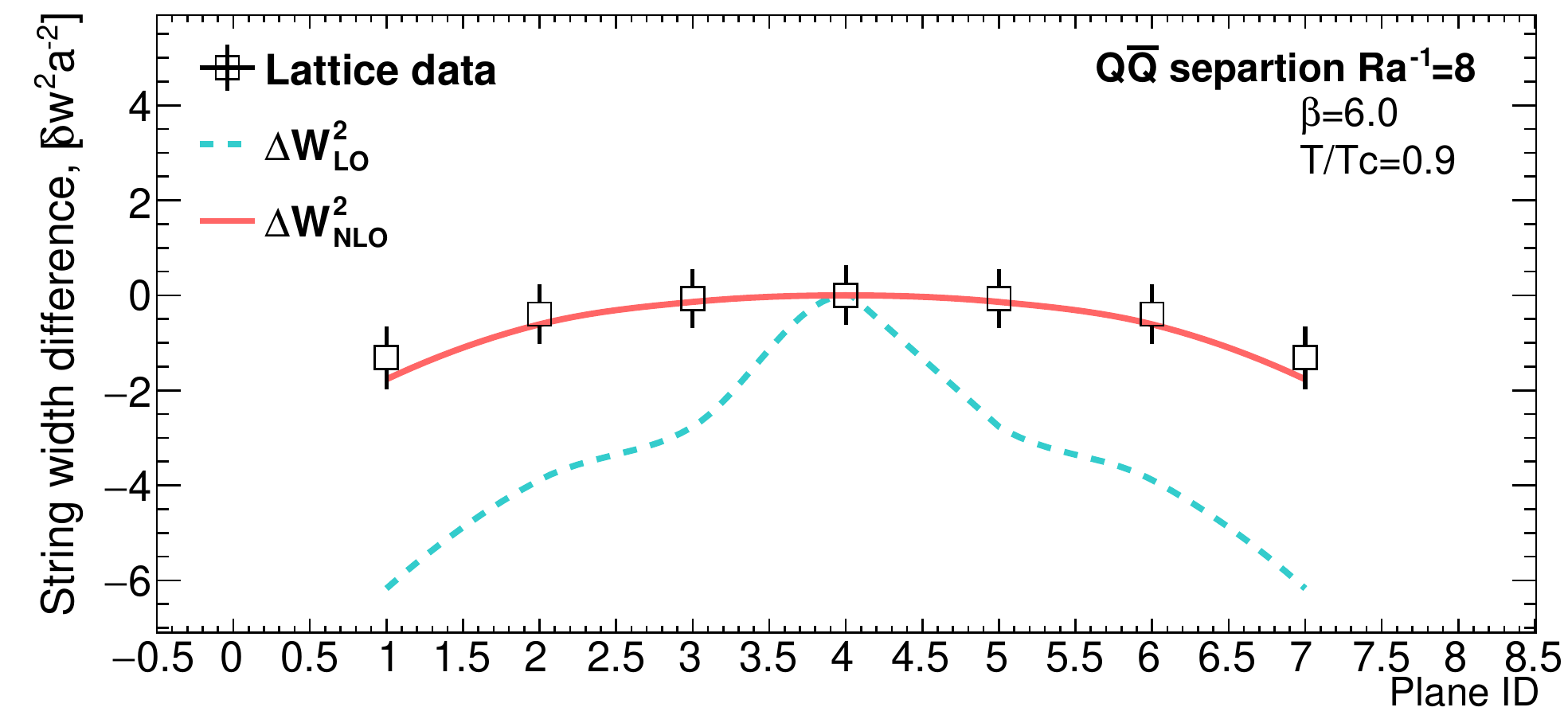}}
\subfigure[$R=0.9$ fm]{\includegraphics[scale=0.4]{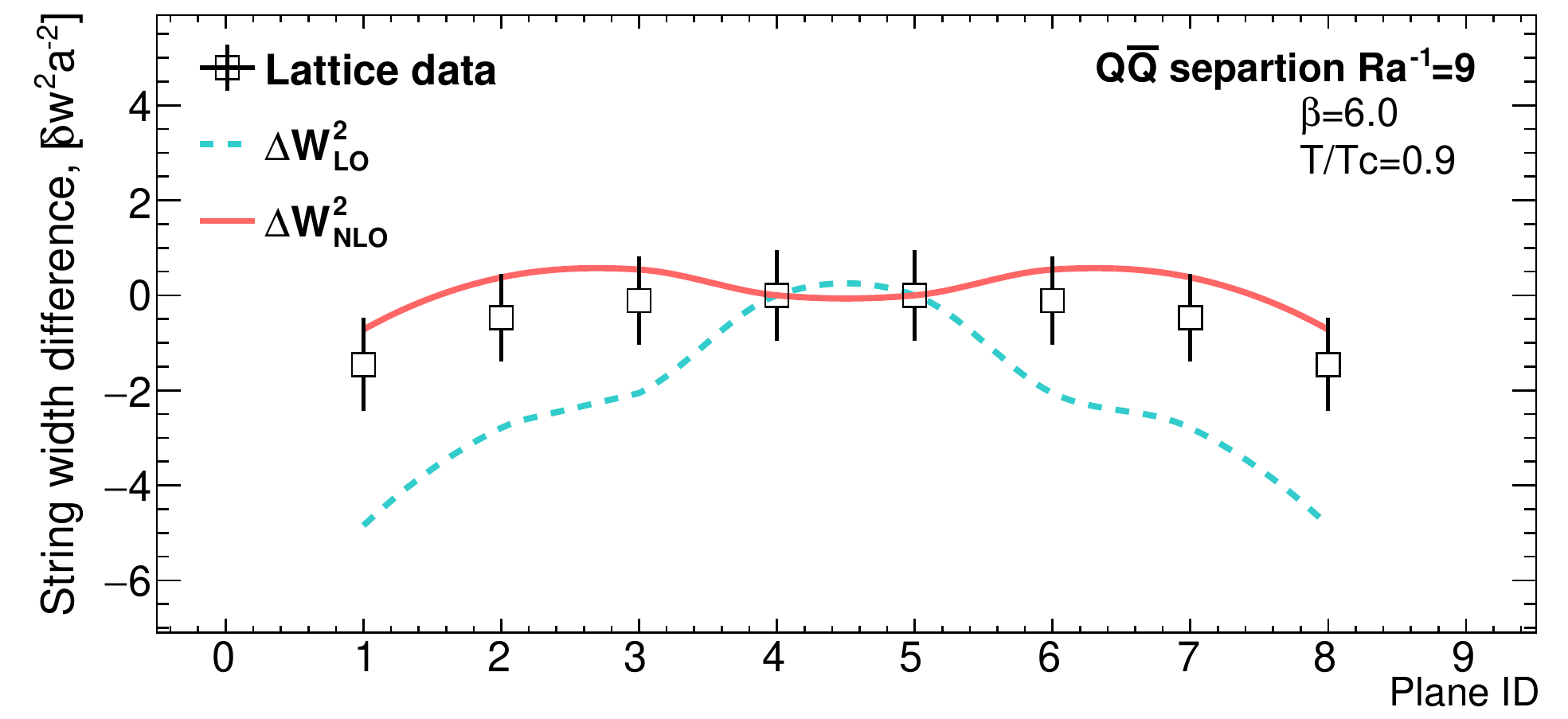}}
\subfigure[$R=1.2$ fm]{\includegraphics[scale=0.4]{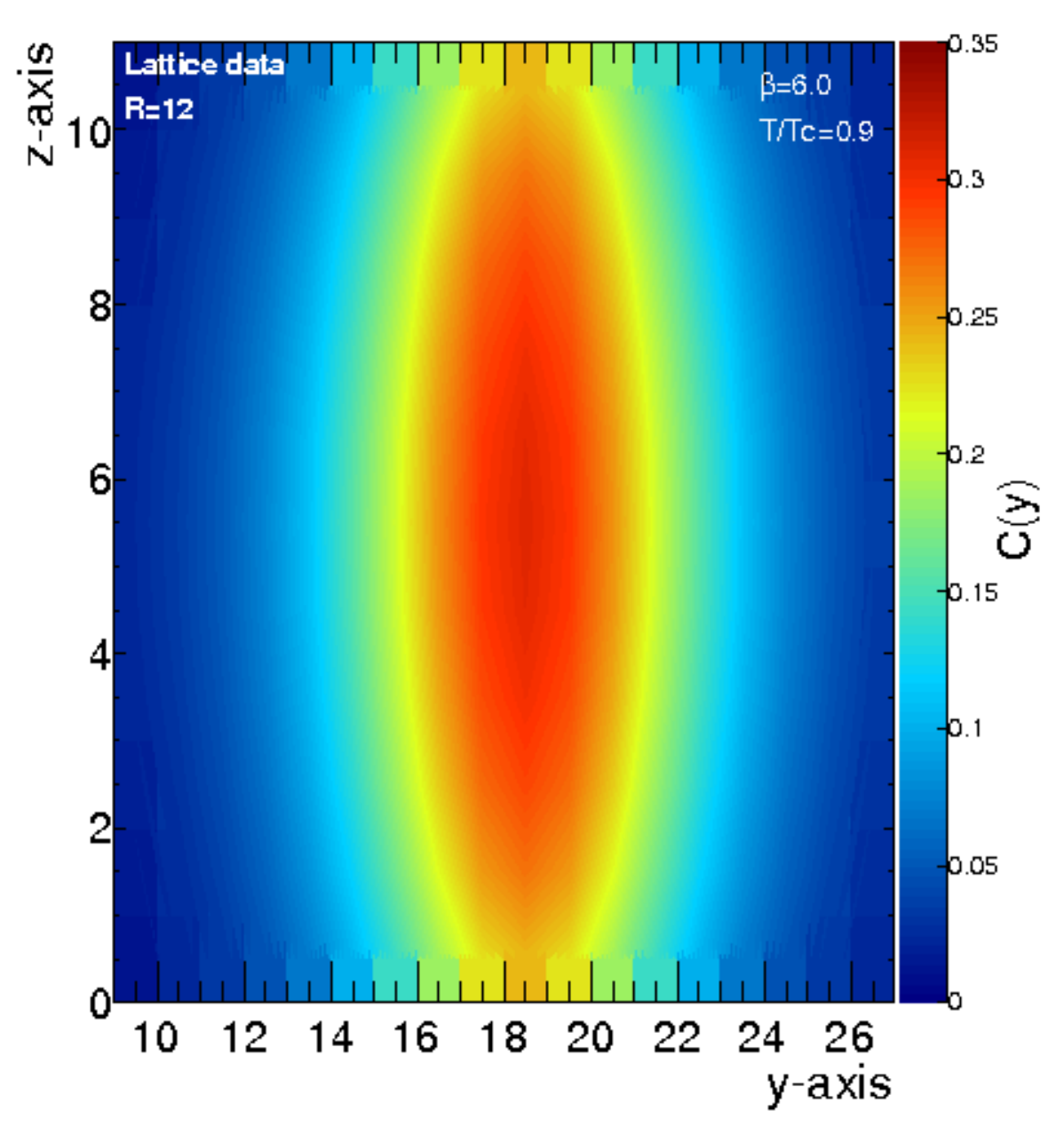}}
\subfigure[$R=1.4$ fm]{\includegraphics[scale=0.4]{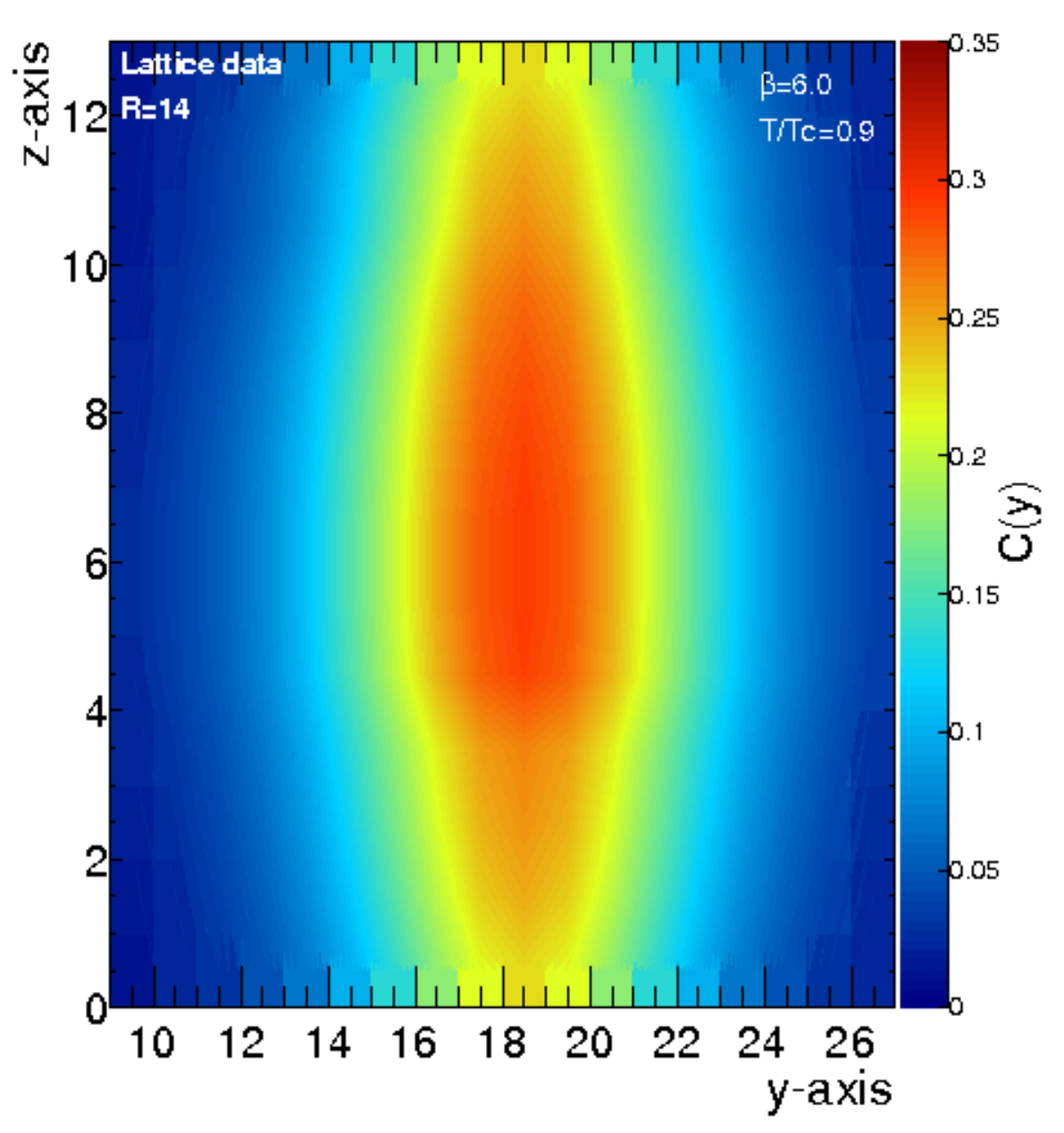}}
\caption{The changes in the width from the middle plane $z=R/2$ at temperature~$ T/T_{x} \approx 0.9$. The co-ordinates $z$ are lattice coordinates (lattice units) and are measured from the quark position $z=0$. The solid and dash lines represent the free-string model (LO) and  self-interacting (NLO) string Eqs.\eqref{WidthNLO} and \eqref{WidthLO}, respectively.}\label{DIFFT09}
\end{center}
\end{figure}     

\begin{equation}
\begin{split}
  W^{2}_{b_4}=&\frac{-\pi^3 (D-2)b_4}{32 R^5 \sigma^2}\bigg(\frac{11 }{36}\text{E}_2(\tau/2)-\frac{5 }{9}\text{E}_2(\tau)-\frac{55}{12}\bigg)\\
  &\left(\,\text{E}_2(\tau)-\frac{5}{4}\right),  
\end{split}
\label{wb4}
\end{equation}

  Equations~\eqref{1wb2} and Eq.~\eqref{wb4} lay out the perturbative expansion of the boundary action and estimate the subsequent augmentation/lessening of mean-square width of the effective string in $D$ dimension at any temperature.

  
\section{Action Density on the Lattice}
\subsection{Width measurements of the Action Density}

 In the following we construct a color-averaged infinitely-heavy static quark-antiquark $Q\bar{Q}$ state by means of two Polyakov lines
\begin{align*}
\mathcal{P}_{2Q}(\vec{r}_{1},\vec{r}_{2}) =  P(\vec{r}_{1})P^{\dagger}(\vec{r}_{2}).
\end{align*}
 
We measure the mean-square width of the action density in SU(3) gluonic configurations. The action density is related to the chromo-electromagnetic fields via $\frac{1}{2}(E^{2}-B^{2})$ and is evaluated via a three-loop improved lattice field-strength tensor~\cite{Bilson}.

  A scalar field characterizing the action density distribution in the Polyakov vacuum or in the presence of color sources~\cite{Bissey} can be defined as 
\begin{equation}
\mathcal{C}(\vec{\rho};\vec{r}_{1},\vec{r}_{2} )= \frac{\langle\mathcal{P}_{2Q}(\vec{r}_{1},\vec{r}_{2}) \, S(\vec{\rho})\,\rangle } {\langle\, \mathcal{P}_{2Q}(\vec{r}_{1},\vec{r}_{2})\,\rangle\, \,\langle S(\vec{\rho})\, \rangle},
\label{Actiondensity}
\end{equation}
  with the vector $\vec{\rho}$ referring to the spatial position of the energy probe with respect to some origin, and the bracket $\langle ...\rangle$ stands for averaging over gauge configurations and lattice symmetries.


  We make use of the symmetry of the four dimensional torus, that is, the measurements taken at a fixed color source's separations $R$ are repeated at each point of the three-dimensional torus and time slice then averaged. The lattice size is sufficiently large to avoid mirror effects or correlations from the other side of the finite size periodic lattice. The characterization Eq.~\eqref{Actiondensity} yields  $C \rightarrow 1$ away from the quarks by virtue of the cluster decomposition of the operators.

\begin{table*}[!hpt]
	\begin{center}
		\begin{ruledtabular}
			\begin{tabular}{ccccccccccc}
				\multirow{1}{*}{Fit} &\multirow{1}{*}{$Q\bar{Q}$ distance,} &\multicolumn{5}{c}{Width of the action density $W_{z}^{2}(x)$}\\	
				\multirow{1}{*}{Range} &\multicolumn{1}{c}{$Ra^{-1}$} &$z=1$ &$z=2$ &$z=3$ &$z=4$ &$z=R/2$\\  				
				\hline
				\multicolumn{7}{c}{$T/T_{c}=0.9$}\\			
				\hline
				\multirow{10}{*}{7--28} 
				&5	&14.1199$\pm$0.098  &13.149$\pm$0.083  &13.149$\pm$0.083  &14.1199$\pm$0.098	&13.149$\pm$0.083\\
				&6	&16.1792$\pm$0.153  &15.223$\pm$0.128  &14.9462$\pm$0.124  &15.223$\pm$0.128	&14.9462$\pm$0.134\\
				&7	&17.96$\pm$0.239  &17.3606$\pm$0.203  &17.2725$\pm$0.203  &17.2725$\pm$0.203	&17.2725$\pm$0.203\\
				&8	&19.3835$\pm$0.363  &19.1211$\pm$0.316  &19.5429$\pm$0.336  &19.8105$\pm$0.358	&19.8105$\pm$0.358\\
				&9	&20.7768$\pm$0.541  &20.5916$\pm$0.477  &21.3668$\pm$0.532  &22.0775$\pm$0.626	&22.0775$\pm$0.626\\
				&10	&22.5249$\pm$0.805  &22.3885$\pm$0.722  &23.2364$\pm$0.817  &23.9204$\pm$1.019	&24.0357$\pm$1.142\\
				&11	&24.1116$\pm$1.166  &24.8579$\pm$1.112  &26.2653$\pm$1.303  &26.6043$\pm$1.618	&25.7155$\pm$1.914\\
				&12	&23.7865$\pm$1.495  &26.9632$\pm$1.649  &29.2494$\pm$0.628  &30.9147$\pm$0.804	&27.2671$\pm$3.295\\
				&13	&21.4501$\pm$1.711  &26.587$\pm$2.094  &31.8025$\pm$0.829  &34.7752$\pm$1.073	&31.3649$\pm$5.516\\
				&14	&19.7353$\pm$2.013  &23.8117$\pm$2.319  &32.5125$\pm$1.067  &38.3333$\pm$1.394	&35.9662$\pm$10.677\\
			
			\end{tabular}		
		\end{ruledtabular}
	\end{center}
	\caption{
		The mean square width of the action density $W^{2}(z)$ measured at the temperature $T/T_{c}=0.9$. The width is estimated with $\sigma_1 \neq \sigma_2$ in Eq.\eqref{conGE} at five consecutive transverse planes $z_i$ to the $Q\bar{Q}$ line.
		}	
	\label{WidthPlanesT09}
\end{table*} 
\begin{table*}[!hpt]
	\begin{center}
		\begin{ruledtabular}
			\begin{tabular}{ccccccccccc}
				\multirow{1}{*}{Fit} &\multirow{1}{*}{$Q\bar{Q}$ distance,} &\multicolumn{5}{c}{Width of the action density $W_{z}^{2}(x)$}\\	
				\multirow{1}{*}{Range} &\multicolumn{1}{c}{$Ra^{-1}$} &$z=1$ &$z=2$ &$z=3$ &$z=4$ &$z=R/2$\\  				
				\hline
				\multicolumn{7}{c}{$T/T_{c}=0.8$}\\				
				\hline
				\multirow{10}{*}{7--28} 
				&5	&7.3822$\pm$0.042  &7.2984$\pm$0.038  &7.3822$\pm$0.042  &7.3822$\pm$0.042	&7.2984$\pm$0.038\\
				&6	&8.2016$\pm$0.068  &8.1932$\pm$0.062  &8.2032$\pm$0.063  &8.1932$\pm$0.062	&8.2032$\pm$0.063\\
				&7	&8.9061$\pm$0.113  &9.0557$\pm$0.105  &9.2114$\pm$0.109  &9.2114$\pm$0.109	&9.2114$\pm$0.109\\
				&8	&9.5545$\pm$0.191  &9.8594$\pm$0.178  &10.2916$\pm$0.192  &10.4928$\pm$0.204	&10.4928$\pm$0.204\\
				&9	&10.3981$\pm$0.321  &10.7443$\pm$0.298  &11.4887$\pm$0.332  &12.1310$\pm$0.378	&12.131$\pm$0.378\\
				&10	&11.8279$\pm$0.553  &11.9383$\pm$0.631  &12.4012$\pm$0.681  &13.7030$\pm$0.853	&14.5621$\pm$0.975\\
				&11	&14.0864$\pm$0.992  &14.0718$\pm$1.201  &13.7877$\pm$1.175  &15.0550$\pm$1.447	&17.2586$\pm$1.909\\
				&12	&17.4755$\pm$1.863  &18.5431$\pm$1.702  &17.0604$\pm$2.460  &18.2364$\pm$1.898	&20.8867$\pm$2.542\\
				&13	&21.2485$\pm$3.256  &24.1686$\pm$3.066  &26.6827$\pm$3.341  &25.0955$\pm$3.766	&28.1915$\pm$5.456\\
				&14	&22.462$\pm$4.675  &26.9322$\pm$4.510  &31.9809$\pm$5.093  &39.9434$\pm$7.260	&50.1043$\pm$16.622\\										
			\end{tabular}		
		\end{ruledtabular}
	\end{center}
	\caption{Similar to Table.~\ref{WidthPlanesT09}, the width of the action density $W_{z}^{2}(x)$ measured at the temperature $T/T_{c}=0.8$ with the use of the fit formula Eq.\eqref{conGE} setting $\sigma_1 \neq \sigma_2$.
		}	
	\label{WidthPlanesT08}
\end{table*}

\begin{table*}[!hptb]
	\begin{center}
		\begin{ruledtabular}
			\begin{tabular}{ccccccccccc}
				\multirow{1}{*}{Fit} &\multirow{1}{*}{$Q\bar{Q}$ distance,} 
				&\multicolumn{2}{c}{Width of the action density $W^{2}(x)$} 
				&\multicolumn{2}{c}{$\chi_{\rm{dof}}^{2}$}
				&\multicolumn{1}{c}{Relative}\\  				
				\multirow{1}{*}{Range} &\multicolumn{1}{c}{$Ra^{-1}$} &$\sigma_1=\sigma_2$ &$\sigma_1\neq \sigma_2$ &$\sigma_1=\sigma_2$& $\sigma_1\neq \sigma_2$ &difference \\  				
	
				\hline
				\multirow{8}{*}{7--28} 
				&5  &10.1524$\pm$0.0563 &13.1490$\pm$0.0826  &63.2042 &0.7365  &22.79\%\\ 				
				&6  &11.5883$\pm$0.0832 &14.9462$\pm$0.1336  &39.2565 &0.2515  &22.47\%\\   				
				&7  &13.4276$\pm$0.1277 &17.2725$\pm$0.2028  &21.7959 &0.0382  &22.26\%\\   				
				&8  &15.4602$\pm$0.1985 &19.8105$\pm$0.3578  &10.2421 &0.0052  &21.96\%\\ 				
				&9  &17.4607$\pm$0.2986 &22.0775$\pm$0.6263  &4.6967  &0.0043  &20.91\%\\   				
				&10 &19.6764$\pm$0.4636 &24.0357$\pm$1.1423  &1.9736  &0.0007  &18.14\%\\ 				
				&11 &21.9126$\pm$0.7022 &25.7155$\pm$1.9140  &1.1621  &0.0094  &14.79\%\\ 				
				&12 &24.3741$\pm$1.1399 &27.2671$\pm$3.2948  &0.8326  &0.0249  &10.61\%\\ 		  											
								
			\end{tabular}		
		\end{ruledtabular}
	\end{center}
	\caption{
		The mean-square width of the action density $W^{2}(z)$ and the corresponding $\chi^{2}$ at the temperature $T/T_{c}=0.9$ in the middle transverse plane intersecting the $Q\bar{Q}$ line  $z=R/2$. The width estimates and the relative differences are obtained in accord to Eq.~\eqref{conGE}, with $\sigma_{1}=\sigma_{2}$ corresponding to the standard Gaussian.}	
	\label{CompWidth}
\end{table*}

  To eliminate statistical fluctuations, uncompromising the physical observable are left intact, only 20 sweeps of  UV filtering using an over-improved algorithm~\cite{Morningstar,Moran} have been applied on all gauge configurations.

  Different UV filtering schemes can be calibrated~\cite{PhysRevD.82.094503,Bonnet} in terms of the corresponding radius of the Brownian motion. The above prescribed number of stout-link sweeps would be the equivalent of 10 sweeps of APE~\cite{Albanese} algorithm~\cite{PhysRevD.82.094503,Bonnet} with an averaging parameter $\alpha=0.7$.

  A careful analysis that we have performed in Ref.~\cite{PhysRevD.82.094503,Bakry:2014ina} ensured that with a number of  $n_{sw}$ of improved cooling sweeps~\cite{Morningstar,Moran} no smearing effects are detectable on either the quark-antiquark $Q\bar{Q}$ potential or the energy density profile for color source separation distances $R \geq 0.5$ fm  which is the distance scale under scrutiny in this investigation.
  
\begin{table*}[!hptb]
	\begin{center}
		\begin{ruledtabular}
			\begin{tabular}{ccccccccccc}
				\multirow{2}{*}{$T/T_{c}=0.9$} &Fit Range &&\multicolumn{5}{c}{$\chi^{2}$ }\\
				&$n=R/a$ &&$z=1$ &$z=2$ &$z=3$ &$z=4$ &$z=R/2$\\
				\hline
				\multirow{9}{*}{\begin{turn}{90}Free String (LO)\end{turn}}				
				&5-9 &&445.232 &425.601 &-- &--    &79.5311\\ 
				&6-9 &&115.849 &149.802 &91.0467 &--    &50.6209\\ 
				&7-9 &&24.3641 &30.7772 &26.5945 &11.6436    &14.2874\\ 				
				&4-12 &&1220.73 &620.346 &1018.86 &--    &84.183\\
				\multicolumn{1}{c}{} &5-12 &&481.859 &467.078 &-- &--    &82.9888\\ 
				\multicolumn{1}{c}{} &6-12 &&137.744 &178.265 &161.094 &39.245    &53.3045\\ 
				\multicolumn{1}{c}{} &7-12 &&36.8014 &48.0687 &78.9119 &38.7756    &15.7182\\ 
				\multicolumn{1}{c}{} &8-12 &&10.9884 &14.6037 &39.1034 &22.4106    &1.9209\\  			
				\multicolumn{1}{c}{} &10-12 &&0.3363 &1.7424 &6.3758 &4.7030 &0.0071\\  								
				\hline
				\multirow{9}{*}{\begin{turn}{90}2 Loops (NLO)\end{turn}}							
				&5-9 &&199.681 &374.824 &-- &--    &79.6823\\ 
				&6-9 &&37.0773 &86.6316 &56.1896 &--    &28.6494\\ 
				&7-9 &&5.4318 &11.6294 &12.0872 &--    &5.4264\\ 								
				&4-12 &&692.334 &1025.6 &615.994 &--    &326.3\\ 
				\multicolumn{1}{c}{} &5-12 &&211.424 &397.556 &-- &--    &80.8514\\ 
				\multicolumn{1}{c}{} &6-12 &&42.9046 &99.4396 &92.2507 &--    &29.2459\\ 
				\multicolumn{1}{c}{} &7-12 &&8.3665  &18.4657 &37.3376 &15.8853   &5.5947\\ 
				\multicolumn{1}{c}{} &8-12 &&2.4583  &5.1658  &18.1117 &10.064    &0.2685\\  
				\multicolumn{1}{c}{} &10-12 &&0.1072 &0.8282  &3.54221 &2.8005    &0.0145\\  
				
			\end{tabular}		
		\end{ruledtabular}
	\end{center}
	\caption{The returned values of $\chi^{2}$ for fit to the free string (LO) Eq.\eqref{WidthLO} and self-interacting NG string (NLO) Eq.~\eqref{WidthNLO} at each selected transverse planes $z_{i}$ at $T/T_{c}=0.9$ with the last column is the retrieved $\chi^{2}$ at the middle of the string $z=R/2$.
	}	
	\label{LO&NLO_fits_PlanesXT09}
\end{table*} 


\begin{table*}[!hpt]
	\begin{center}
		\begin{ruledtabular}
			\begin{tabular}{ccccccccccc}
				\multirow{2}{*}{$T/T_{c}=0.8$} &Fit Range &&\multicolumn{5}{c}{$\chi^{2}$ }\\
				&$n=R/a$ &&$z=1$ &$z=2$ &$z=3$ &$z=4$ &mid.plane\\
				\hline
				\multirow{7}{*}{\begin{turn}{90}Free String(LO)\end{turn}}				
				&4-12 &&578.806 &453430 &-- &--    &880374\\ 
				\multicolumn{1}{c}{} &5-12 &&60.5188 &36402.6 &266299 &--    &168150\\ 
				\multicolumn{1}{c}{} &6-12 &&44.1409 &31.0112 &126.596 &--  &189.559\\ 
				\multicolumn{1}{c}{} &7-12 &&19.8978 &10.4049 &4.0910 &3.0459    &2.7246\\ 
				\multicolumn{1}{c}{} &8-12 &&10.7254 &5.3255 &2.2323 &2.6146    &2.7243\\ 
				\multicolumn{1}{c}{} &10-12 &&2.5551 &2.0697 &0.7103 &0.6643  &0.6747\\
				\\ 							
				\hline
				\multirow{6}{*}{\begin{turn}{90}2 Loops(NLO)\end{turn}}	
				&4-12 &&241.135 &157140 &-- &--    &69091.1\\ 
				\multicolumn{1}{c}{} &5-12 &&93.9843 &7415.33 &-- &--  &9118.51\\ 
				\multicolumn{1}{c}{} &6-12 &&37.3147 &18.86 &26.4692 &--   &53.3679\\ 
				\multicolumn{1}{c}{} &7-12 &&12.6149 &14.91 &9.0537 &2.8830    &4.1864\\ 
				\multicolumn{1}{c}{} &8-12 &&7.0765 &4.6559 &2.1293 &2.5598   &2.6349\\ 
				\multicolumn{1}{c}{} &10-12 &&2.0170 &1.8054 &0.5687 &0.5212 &0.5480\\ 				
			\end{tabular}		
		\end{ruledtabular}
	\end{center}
	\caption{Enlisted are the returned value of $\chi^{2}$ corresponding to the fit to  Nambu-Goto string in the leading-order (LO) Eq.\eqref{WidthLO} and the next to leading order (NLO) Eq.\eqref{WidthNLO} formulation at each selected transverse planes $z_{i}$, the last column corresponds to resultant fit at the middle plane of the string $z=R/2$.}	
	\label{LO&NLO_fits_PlanesXT08}
\end{table*}

  To estimate the mean-square width of the gluonic action density a long each transverse plane to the quark-antiquark axis. Taking into consideration the axial cylindrical symmetry of the tube, we choose a double Gaussian function of the same amplitude, $A$, and mean value $\mu=0$

\begin{equation}  
G(r,\theta;z)= A (e^{-r^2/\sigma_1^2}+e^{-r^2/\sigma_2^2})+\kappa,
\label{conGE}
\end{equation}

  In the above form the constraint  $\sigma_1=\sigma_2$ corresponds to the standard Gaussian distribution. Table.~\ref{CompWidth} compares the returned value of the $\chi^{2}$ for both optimization ansatz, namely, the constrained form $\sigma_1=\sigma_2$  and  $\sigma_1 \neq \sigma_2$ unconstrained form. The fits of the double Gaussian form return acceptable values of $\chi^{2}$ at the intermediate distances.

   Good $\chi^{2}$ values are returned as well when fitting the action density profile to a convolution of the Gaussian with an exponential~\cite{PhysRevD.88.054504,Bakry:2014gea}, however, considering statistical uncertainties at large distances (see Fig.~\eqref{action}) we opt to the form Eq.~\ref{conGE} with $\sigma_1 \neq \sigma_2$ for stable fits.
 
   A measurement of the width of the string's action density may be taken by fitting the density distribution $\mathcal{C}(\vec{\rho};z)$ to Eq.~\eqref{conGE} through each transverse to the cylinder's axis $z$ to Eq.~\eqref{conGE}
\begin{equation}
\label{width1}
     \mathcal{C}(r,\theta;z)=1-G(r,\theta;z)
\end{equation} 
 \noindent  with $r^2=x^2+y^2$ in each selected transverse plane $\vec{\rho}(r,\theta;z)$. The second moment of the action density distribution with respect to the cylinder's axis $z$ joining the two quarks is 

\begin{equation} 
\label{widthg}
W^{2}(z)=\quad \dfrac{\int \, dr\,r^{3}\,G(r,\theta;z)} {\int \,dr \,r \: G(r,\theta;z) },
\end{equation}  

  which defines the mean square width of the tube on the lattice. The locus of the color sources corresponds to $z=0$ or $z=R$, respectively.

  In Table~\ref{CompWidth} the numerical values of the mean-square width of the string at the middle plane between the two color sources are in-listed. The percentage differences in the measured width measured with the use of both ansatz in Table~\ref{CompWidth} indicate an almost constant shift amounting approximately to $22\%$ of that measured using unconstrained optimization $\sigma_{1} \neq \sigma_{2}$. 

   Further measurements of the mean-square width at consecutive transverse planes $z=1$ to $z=4$ are enlisted in Table~\ref{WidthPlanesT09} of Appendix.~A. The width is estimated in accord to Eqs.\eqref{conGE} and \eqref{widthg} at each selected plane $z_i$ fixed with respect to one color source. We found the unconstrained optimization Eq.\eqref{conGE} is returning $\sigma_{1} \neq \sigma_{2}$ at all color separation distances.

   The numerical values in Table~\ref{WidthPlanesT09} are indicating a broadening in mean-square width of the string at all transverse planes $z_{i}$ as the color sources are pulled apart. The plot of the width at consecutive planes in Fig.~\ref{PlanesID1234T09} more clearly depicts an increasing slop in the pattern of growth as one considers farther planes from the quark sources up to a maximum slop in the middle plane.
   
\subsection{Pure Nambu-Goto String}
   The broadening of the width at each selected transverse plane can be compared to that of the corresponding width of the quantum string Eqs.\eqref{WidthLO} and \eqref{WidthNLO}. Our discussion in Ref.~\cite{Bakry:2018kpn, Bakry:2017utr} for the fit analysis of the two Polyakov loop correlator enlightens that both the LO and NLO approximations are substantial different when the temperature $T/T_{c}\simeq0.9$ is close to the deconfinement point.
   
\begin{table*}[!hpt]
	\begin{center}
		\begin{ruledtabular}
			\begin{tabular}{ll|ccc|ccc|ccc}
				\multirow{2}{*}{} &~~~~~~~~~~~~Fit Range 
				&\multicolumn{3}{c}{$R \in[0.4,0.7]$ fm} 
				&\multicolumn{3}{c}{$R \in[0.5,0.8]$ fm}
				&\multicolumn{3}{c}{$R \in[0.6,0.9]$ fm}\\
				\multicolumn{2}{c|}{String Type} 
				&\multicolumn{1}{c}{$\chi^{2}$} &\multicolumn{1}{c}{$R_{0}$} 
				&\multicolumn{1}{c}{$\alpha$} &\multicolumn{1}{c}{$\chi^2$}&\multicolumn{1}{c}{$R_{0}$}&\multicolumn{1}{c}{$\alpha$}&\multicolumn{1}{c}{$\chi^2$}&\multicolumn{1}{c}{$R_{0}$}&\multicolumn{1}{c}{$\alpha$}\\
				\hline				
\multirow{2}{*}{\begin{turn}{0} {\scriptsize{$z=1$}} \end{turn}}	 
 &\small{$NG_{{\ell o}}+PK_{{\ell o}}$} &42.58&2.3(1)&0.14(2)&0.45&3.3(2)&0.20(5)&0.33&3.2(2)&0.1(1)       \\
\multicolumn{1}{c}{} &\small{$NG_{{n\ell o}}+PK_{{\ell o}}$}&21.90&1.6(2)&0.086(7)&3.46&2.6(3)&0.04(3)&1.74&2.0(3)&0.18(9) \\\hline
\multirow{2}{*}{\begin{turn}{0} {\scriptsize{$z=2$}} \end{turn}}	 
 &\small{$NG_{{\ell o}}+PK_{{\ell o}}$} &96.69&1.3(1)&0.17(2)&1.11&2.7(2)&0.30(4)&0.53&2.8(2)&0.36(9)       \\
\multicolumn{1}{c}{} &\small{$NG_{{n\ell o}}+PK_{{\ell o}}$}&49.20&0.2(2)&0.029(6)&0.92&1.7(2)&0.21(3)&2.67&1.3(3)&0.11(8) \\\hline
\multirow{2}{*}{\begin{turn}{0} {\scriptsize{$z=3$}} \end{turn}}	 
 &\small{$NG_{{\ell o}}+PK_{{\ell o}}$} &146.38&0.7(1)&0.33(2)&9.70&2.6(2)&0.28(4)& 0.16&3.1(2)&0.6(1)      \\
\multicolumn{1}{c}{} &\small{$NG_{{n\ell o}}+PK_{{\ell o}}$}&79.11&0.8(2)&0.043(6)&2.82&1.5(3)&0.21(3)&0.58&1.7(3)&0.33(8)\\\hline
\multirow{2}{*}{\begin{turn}{0} {\scriptsize{$z=4$}} \end{turn}}	 
 &\small{$NG_{{\ell o}}+PK_{{\ell o}}$} &&&&2.07&2.9(2)&0.33(5)&2.72&3.2(2)&0.6(1)       \\
\multicolumn{1}{c}{} &\small{$NG_{{n\ell o}}+PK_{{\ell o}}$}&&&&0.98&1.9(3)&0.23(3)&0.89&2.0(3)&0.29(8) \\\hline
\multirow{2}{*}{\begin{turn}{0} {\scriptsize{$z=\frac{R}{2}$}} \end{turn}}	 
 &\small{$NG_{{\ell o}}+PK_{{\ell o}}$} &84.42&1.1(1)&0.19(2)&12.41&2.7(2)&0.29(4)&1.30&3.2(2)&0.7(1)\\
\multicolumn{1}{c}{} &\small{$NG_{{n\ell o}}+PK_{{\ell o}}$}&38.15&0.1(2)&0.032(6)&4.40&1.6(3)&0.21(3)&0.22&2.1(3)&0.41(9)\\              
			\end{tabular}		
		\end{ruledtabular}
	\end{center}
	\caption{The returned $\chi^2$ from the fits of the leading order NG string width Eq.\eqref{WidthLO}, next-to-leading width of NG string Eq.\eqref{WidthNLO} and combination of rigidity contributions given by PK string models Eqs.\eqref{WExt}. The Monte-Carlo lattice data of the square-width of the action density width.}
\label{Fits1NGR}	
\end{table*}  

\begin{table*}[!hpt]
	\begin{center}
		\begin{ruledtabular}
			\begin{tabular}{ll|ccc|ccc|ccc}
				\multirow{2}{*}{} &\scriptsize{Plane/Fit Interval} 
				&\multicolumn{3}{c}{$R \in[0.5,1.2]$ fm} 
				&\multicolumn{3}{c}{$R \in[0.6,1.2]$ fm}
				&\multicolumn{3}{c}{$R \in[0.7,12]$ fm}\\
				\multicolumn{2}{c|}{\scriptsize{String Type}} 
				&\multicolumn{1}{c}{$\chi^{2}$} &\multicolumn{1}{c}{$R_{0}$} &\multicolumn{1}{c}{$\alpha$} 
				&\multicolumn{1}{c}{$\chi^2$} &\multicolumn{1}{c}{$R_{0}$} &\multicolumn{1}{c}{$\alpha$}
				&\multicolumn{1}{c}{$\chi^2$} &\multicolumn{1}{c}{$R_{0}$} &\multicolumn{1}{c}{$\alpha$}\\
				\hline 
\multirow{2}{*}{\begin{turn}{0} {\scriptsize{$z=1$}} \end{turn}}	 
 &\small{$NG_{{\ell o}}+PK_{{\ell o}}$} &2.46&3.2(2)&0.18(4)&1.46&3.2(2)&0.1(9)&1.01&3.2(2)&0.0(2)
    \\
\multicolumn{1}{c}{} &\small{$NG_{{n\ell o}}+PK_{{\ell o}}$}&18.58&2.1(3)&0.01(2)&6.37&1.7(3)&0.26(8)&2.06&1.9(3)&0.6(2)\\\hline
\multirow{2}{*}{\begin{turn}{0} {\scriptsize{$z=2$}} \end{turn}}	 
 &\small{$NG_{{\ell o}}+PK_{{\ell o}}$} &2.69&2.8(1)&0.31(4)&1.50&2.8(1)&0.38(8)&1.43&2.86(2)&0.3(2)   \\
\multicolumn{1}{c}{} &\small{$NG_{{n\ell o}}+PK_{{\ell o}}$}&7.06&1.4(2)&0.19(2)&3.83&1.2(2)&0.07(7)&0.57&1.3(3)&0.2(2) \\\hline
\multirow{2}{*}{\begin{turn}{0} {\scriptsize{$z=3$}} \end{turn}}	 
&\small{$NG_{{\ell o}}+PK_{{\ell o}}$} &44.90&3.1(1)&0.40(4)&12.02&3.4(1)&0.78(7)&7.25&3.2(2)&1.0(1) \\
\multicolumn{1}{c}{} &\small{$NG_{{n\ell o}}+PK_{{\ell o}}$}&7.8&1.6(2)&0.22(2)&3.10&1.8(2)&0.36(7)&3.09&1.8(3)&0.4(1)\\\hline
\multirow{2}{*}{\begin{turn}{0} {\scriptsize{$z=4$}} \end{turn}}	 
 &\small{$NG_{{\ell o}}+PK_{{\ell o}}$}& 45.93&3.3(1)&0.45(4)&22.56&3.6(2)&0.80(8)&7.54&3.4(2)&1.3(2)\\
\multicolumn{1}{c}{} &\small{$NG_{{n\ell o}}+PK_{{\ell o}}$}&11.76&2.1(2)&0.25(3)&8.19&2.4(3)&0.38(7)&3.83&2.2(3)&0.7(2)
 \\\hline
\multirow{2}{*}{\begin{turn}{0} {\scriptsize{$z=\frac{R}{2}$}} \end{turn}}	 
 &\small{$NG_{{\ell o}}+PK_{{\ell o}}$} &20.03&2.9(2)&0.34(4)&1.44&3.3(2)&0.7(1)&0.84&3.3(2)&0.9(2)\\
\multicolumn{1}{c}{} &\small{$NG_{{n\ell o}}+PK_{{\ell o}}$}&5.45&1.6(2)&0.22(2)&0.69&2.0(3)&0.39(8)&0.69&2.0(3)&0.4(2)\\
			\end{tabular}		
		\end{ruledtabular}
	\end{center}
	\caption{Same as Table~\ref{Fits1NGR}; however, the fit range includes larger color source separations.}
\label{Fits2NGR}	
\end{table*}  
 
  In Table~\ref{LO&NLO_fits_PlanesXT09} summarized are the resultant values of the fit considering various range of sources’ separations. For this fit procedure the string tension is fixed to its value returned  at $T/T_{c}\simeq 0.8$ form fits of $Q\bar{Q}$ data.

The free string (LO) Eqs.\eqref{WidthLO} and self-interacting Eq.~\eqref{WidthNLO} (NLO) solutions are  one parameter fit functions in the ultraviolet cutoff $R(\xi)$. While in the LO formula the ultraviolet cutoff has the effect of a constant shift in the flux-tube width, the UV cutoff alters the slop in the NLO formula of NG string.  

  The leading order approximation would show a strong dependency on the fit range if the data points at small sources’ separations are considered. The first three entries in Table~\ref{LO&NLO_fits_PlanesXT09} compares the value of $\chi^{2}$ for both approximations at source separations $R=0.5$ fm up to $R=0.9$ fm, that is, excluding the last three points. The free string picture poorly describes the lattice data at short distances.

  With the data points at short distances excluded from the fit the values of $\chi^{2}$ decrease gradually. For example, first four points excluded from the fit, the returned $\chi^2$ is smaller, indicating that only the data points at large source separation are parameterized by the string model formula. With the consideration of the next leading order solution of the NG action the values of $\chi^{2}$ are reduced. Nevertheless, the values of  $\chi^{2}$ are still significantly too large to precisely match the numerical data in intermediate distances.

  Large values of $\chi^{2}$ are retrieved if a source separation such as $R=4a$ to $R=12a$ is included for either LO and NLO approximations. This is, eventhough, a relative improvements when considering the the two loop approximation. The fits  in Table~\ref{LO&NLO_fits_PlanesXT09} divulge a strong dependency on the fit range if the points at small sources’ separations are excluded with a gradual decrease in the values of $\chi^{2}$ and stablity in the fits.

  In Figs.~\ref{MT09} and ~\ref{PlanesID1234T09} are plots of the fit for the mean-square width at the middle plane of the tube, $z=R/2$ together with the corresponding fits to the free string Eqs.\eqref{WidthLO} and the self-interacting NLO form Eq.\eqref{WidthNLO}. The fit range for the free string  Eqs.\eqref{WidthLO} is chosen for color source separations extending from $R=1.0$ fm to $R=1.2$ fm, however, for the  NLO self interacting Eq.\eqref{WidthNLO} string, the fit range includes two additional points $R=0.8-1.2$ fm. The fit regions in Figs~\ref{MT09} and Fig.~\ref{PlanesID1234T09} are chosen so that both approximations give almost the same behavior in the asymptotic region at large color source separations $R \geq 10a$. Thus, in order to approach the NLO approximation in the asymptotic region fits to the leading order approximation Eqs.\eqref{WidthLO} should on considered on large distances.  
 
\begin{table*}[!hpt]
	\begin{center}
		\begin{ruledtabular}
			\begin{tabular}{ll|ccc|ccc|ccc}
				\multirow{2}{*}{$T/T_c=0.9$} &\scriptsize{Plane/Fit Range} 
				&\multicolumn{3}{c}{$R \in[0.4,0.7]$ fm} 
				&\multicolumn{3}{c}{$R \in[0.5,0.8]$ fm}
				&\multicolumn{3}{c}{$R \in[0.6,0.9]$ fm}\\
				\multicolumn{2}{c|}{\scriptsize{String Type}} 
				&\multicolumn{1}{c}{$\chi^{2}$} &\multicolumn{1}{c}{$R_{0}$} 
				&\multicolumn{1}{c}{$b_2$} &\multicolumn{1}{c}{$\chi^2$}&\multicolumn{1}{c}{$R_{0}$}&\multicolumn{1}{c}{$b_2$}&\multicolumn{1}{c}{$\chi^2$}&\multicolumn{1}{c}{$R_{0}$}&\multicolumn{1}{c}{$b_2$}\\
				\hline				
                                \multirow{2}{*}{\begin{turn}{0} {\scriptsize{$z=1$}} \end{turn}}	 
                                 &\small{$NG_{{\ell o}}+b_2$}&   143.9 &8.8(1)&2.7(2)&11.9&12.6(3)&11.3(6)&2.7&15.1(6)&19.3(1.8)\\
                                \multicolumn{1}{c}{} &\small{$NG_{{n\ell o}}+b_2$}&67.97&10.6(2)&-4.11(5)&1.3&13.6(4)&-1.5(3)&0.056&14.3(7)&-0.31(1.0)\\\hline
\multirow{2}{*}{\begin{turn}{0} {\scriptsize{$z=2$}} \end{turn}}	 
 &\small{$NG_{{\ell o}}+b_2$} &216.3&2.8(1)&0.08(0.14)&27.2&6.9(2)&9.0(5)&3.2&9.9(5)&18.9(1.5)\\
\multicolumn{1}{c}{} &\small{$NG_{{n\ell o}}+b_2$}&155.3&2.5(2)&-0.16(4)&9.6&6.6(3)&3.2(2)&0.2&8.3(6)&6.2(9)\\\hline
\multirow{2}{*}{\begin{turn}{0} {\scriptsize{$z=3$}} \end{turn}}	 
 &\small{$NG_{{\ell o}}+b_2$}&225.0&-0.9(1)&-6.6(1)&39.7&3.4(3)&2.7(5)&5.8&7.1(5)&14.9(1.7)\\
\multicolumn{1}{c}{} &\small{$NG_{{n\ell o}}+b_2$}&199.6&-2.5(1)&-2.69(5)&21.3&2.3(3)&1.2(2)&0.97&5.1(6)&6.1(9) \\\hline
\multirow{2}{*}{\begin{turn}{0} {\scriptsize{$z=4$}} \end{turn}}	 
 &\small{$NG_{{\ell o}}+b_2$} &---&---&---&28.9 &0.6(3)&-8.5(6)&10.32&3.9(5)&1.9(1.7)\\
\multicolumn{1}{c}{} &\small{$NG_{{n\ell o}}+b_2$}&---&---&---&15.8&-1.0(3)&-6.7(3)&3.7&1.5(6)&-2.4(1.0)\\\hline
\multirow{2}{*}{\begin{turn}{0} {\scriptsize{$z=\frac{R}{2}$}} \end{turn}}	 
 &\small{$NG_{{\ell o}}+b_2$}&64.3&1.1(1)&-1.7(1)&17.97&3.4(3)&3.4(5)&3.2&5.9(5)&11.6(1.7)\\
\multicolumn{1}{c}{} &\small{$NG_{{n\ell o}}+b_2$}&41.23&0.2(2)&0.04(4)&6.83&2.4(3)&1.8(2)& 0.62&4.1(6)&4.7(1.0)\\
			\end{tabular}		
		\end{ruledtabular}
	\end{center}
	\caption{The returned $\chi^2$ from the fits of either the leading order NG string width Eq.~\eqref{WidthLO} or next-to-leading width of NG string Eq.~\eqref{WidthNLO} together with the thickness terms due to the boundary action $W^2_{b_2}$ given by models Eq.~\eqref{Wid_LO_Boundb2} and Eq.~\eqref{Wid_NLO_Boundb2}. The Monte-Carlo lattice data of the square-width of the action density width.}
\label{Fits1NGb2}	
\end{table*}

\begin{table*}[!hpt]
	\begin{center}
		\begin{ruledtabular}
			\begin{tabular}{ll|ccc|ccc|ccc}
				\multirow{2}{*}{$T/T_c=0.9$} & \scriptsize{Plane/Fit Interval} 
				&\multicolumn{3}{c}{$R \in[0.4,1.2]$ fm} 
				&\multicolumn{3}{c}{$R \in[0.5,1.2]$ fm}
				&\multicolumn{3}{c}{$R \in[0.6,1.2]$ fm}\\
				\multicolumn{2}{c|}{\scriptsize{String Type}} 
				&\multicolumn{1}{c}{$\chi^{2}$} &\multicolumn{1}{c}{$R_{0}$} &\multicolumn{1}{c}{$b_2$} 
				&\multicolumn{1}{c}{$\chi^2$} &\multicolumn{1}{c}{$R_{0}$} &\multicolumn{1}{c}{$b_2$}
				&\multicolumn{1}{c}{$\chi^2$} &\multicolumn{1}{c}{$R_{0}$} &\multicolumn{1}{c}{$b_2$}\\
				\hline 
\multirow{2}{*}{\begin{turn}{0} {\scriptsize{$z=1$}} \end{turn}}	 
&\small{$NG_{{\ell o}}+b_2$}&355.4&9.6(1)&3.6(1)&52.6&13.5(3)&12.8(6)&14.98&16.1(5)&22.2(1.6)\\
\multicolumn{1}{c}{} &\small{$NG_{{n\ell o}}+b_2$}&117.9&11.2(2)&-4.14(5)&5.7&13.9(3)&-1.4(3)&2.4&14.8(6)&0.4(1.0)\\\hline
\multirow{2}{*}{\begin{turn}{0} {\scriptsize{$z=2$}} \end{turn}}	 
&\small{$NG_{{\ell o}}+b_2$}&538.6&3.6(1)&0.9(1)&216.3&25.9&10.9(4)&25.9&10.9(4)&21.7(1.4)\\
\multicolumn{1}{c}{}&\small{$NG_{{n\ell o}}+b_2$}&296.8&3.4(2)&-0.19(5)&27.5&7.1(3)&3.4(2)&7.9&8.9(5)&7.1(9)\\\hline
\multirow{2}{*}{\begin{turn}{0} {\scriptsize{$z=3$}} \end{turn}}	 
&\small{$NG_{{\ell o}}+b_2$} &851.9&0.2(1)&-5.4(2)&851.9&4.9(2)&5.6(5)&96.3&9.6(4)&22.47(1.4)\\
\multicolumn{1}{c}{} &\small{$NG_{{n\ell o}}+b_2$}&566.8&-1.2(2)&-2.72(5)&118.4&3.5(3)&1.9(2)&42.4&7.1(5)&9.0(1)\\\hline
\multirow{2}{*}{\begin{turn}{0} {\scriptsize{$z=4$}} \end{turn}}	 
&\small{$NG_{{\ell o}}+b_2$}&--- &--- &---&194.95&1.8(2)&-6.1(5)&88.9&6.0(5)&8.4(1.5)\\
\multicolumn{1}{c}{} &\small{$NG_{{n\ell o}}+b_2$}&--- &--- &--- &98.7 &-0.007(0.3)&-6.0(3)&44.9&3.3(5)&0.3(9)\\\hline
\multirow{2}{*}{\begin{turn}{0} {\scriptsize{$z=\frac{R}{2}$}} \end{turn}}	 
&\small{$NG_{{\ell o}}+b_2$}&147.2&1.4(1)&-1.3(1)&31.6&3.7(2)&4.0(5)&4.3&6.1(5)&12.2(1.6)\\
\multicolumn{1}{c}{} &\small{$NG_{{n\ell o}}+b_2$}&74.7&0.5(1)&0.02(5)&9.1&2.5(3)&1.9(2)&0.77&4.0(6)&4.6(1.0)\\
			\end{tabular}		
		\end{ruledtabular}
	\end{center}
	\caption{Same as Table~\ref{Fits1NGb2}; however, the fit range includes larger color source separations.}
\label{Fits2NGb2}		
\end{table*}

  The string fluctuations have an almost constant cross-section at the intermediate distances $0.8 \textless R \textless 1.1$ fm which is not what is expected from the free string approximation  Eqs.\eqref{WidthLO}~\cite{allais,PhysRevD.82.094503,Bakry:2010sp} at this distance scale. The analysis of the lattice data has revealed curvatures along the planes transverse to the quark-antiquark line at large distances~\cite{PhysRevD.82.094503,Bakry:2010sp}. In the intermediate distances the profile along the transverse planes is geometrically more flat than the free-string picture would imply.


  Re-render of the mean-square width of lattice data together with fits to Eqs.\eqref{WidthLO} and \eqref{WidthNLO} discloses the geometrical effects of the inclusion of NLO order terms. The width $W^2(z) a^{-2}$ at the middle plan $z=R/2$ is subtracted from that at the plane $W^2(z_i)$, shown in Fig~\ref{DIFFT09} for two typical  $Q\bar{Q}$ configurations at $R=0.8$ fm and $R=0.9$ fm. The string profile when considering the NLO terms in the effective action Eq.\eqref{WidthNLO} show improvements in the match with lattice data. The suppression of the tube curvature  and the constant width property at the intermediate region can be conceived as geometrical features due to the higher loops in the string interactions.   

  Although the statistical fluctuations increase with the decrease of the temperature (see Fig.~\ref{HIST0809}), the width estimates obtained through fitting action density to Eq.\eqref{conGE} can be stabilized with the use of the standard Gaussian form $\sigma_{1}=\sigma_{2}$ in Eq.\eqref{conGE} at the temperature $T/T_c=0.8$ instead.
  
   
    Our expectations from the fit behavior of $Q\bar{Q}$ potential at $T/T_{c}=0.8$ to both the LO and NLO formulas that higher order effects are negligible at this temperature scale.
   
    Most of the considerations concerning the validity of both approximations to the $Q\bar{Q}$ potential at the temperature $T/T_{c}=0.8$ seem to hold as well for the string profile. Considering the same fit range, the solid and dashed lines corresponding to the two approximations (LO) Eq.\eqref{WidthLO} and (NLO) Eq.\eqref{WidthNLO} in Fig.~\ref{MT08} almost coincide, with exception of subtlety at the end point $R=0.5$ fm. The mismatch at 0.5 fm is less obvious when considering fits at other transverse planes than the middle as can be seen in Fig~\ref{PlanesID1234}. This can be attributed to the high value of $\chi^{2}_{dof}$ (only at $R=0.5$ and $R=0.6$ fm) when measuring the width through the standard Gaussian distribution, i.e, $\sigma_1=\sigma_2$ in Eq.~\eqref{conGE}. We report numerical analysis with other fit functions that depends on the separation between the quarks in a next version.

    At temperature  $T/T_c=0.8$, the fit results summarized in Tables~\ref{LO&NLO_fits_PlanesXT08} of the LO and NLO forms of the NG string return very close parameterization behavior in both the asymptotic and intermediate distances regions regardless of the selected fit range. Indeed, higher order effects are almost suppressed at this temperature scale.
    


  In Table~\ref{LO&NLO_fits_PlanesXT08} the fit to the LO approximation unveils good values of $\chi^2$  for color source separation up to $R=0.6$ fm, the next to leading order fits, however, improves with respect to fit range when including these source separations $R=0.5$ fm and $R=0.6$ fm, this manifests at the middle and other the consecutive planes $z$ as well.

  At the same temperature $T/T_{c}=0.8$, a rendering compares the width difference $\delta W^2=\vert W^2(z_i)-W^{2}(R/2)\vert$ of lattice data and the corresponding fits to string model in Fig.~\ref{DIFFT08}. The display in the figure width at each plane is subtracted from the middle plan $z=R/2$. This unveils an almost constant width along the transverse planes to the color sources. The curvatures induced by thermal effects~\cite{Bakry:2010sp, PhysRevD.82.094503} only manifests at temperatures closer to the deconfinement point and at large distances. This shows that regardless of the diminish of the thermal form factors the flux-tube density lines assumes the same shape.
  
\subsection{Polyakov-Kleinert/Rigid String}  
  Our expectations based on the fit analysis of the $Q\bar{Q}$ potential data of the ordinary Nambu-Goto string Eq.~\eqref{NGLO}, Eq.\eqref{NGNLO} and Polyakov-Kleinert action Eq.\eqref{S_extlo} are substantial improvements at the temperature very close to the deconfinement point  $T/T_{c}\simeq0.9$. These improvements have been observed in the compact Abelian $U(1)$ gauge model as well~\cite{Caselle:2016mqu}.

  Let us again set the broadening of the width at each selected transverse plane into comparison with, however, that of the rigid/smooth strings Eqs.\eqref{WExt} and Eq.\eqref{Ext2loopT}. We summarize the resultant $\chi^{2}$ of the fits in Table.~\ref{Fits1NGR} and Table.~\ref{Fits2NGR}. In the first entries the returned parameter is the renormalization $R_{0}$ for the fit to the width at the leading order and the next-to-leading order terms of NG string Eq.~\eqref{WidthLO} and Eq.~\eqref{WidthNLO}, respectively. The following entries are for the rigidity $\alpha$ parameters from the fits to the mean-square width of the relevant order for both NG and PK string Eqs.~\eqref{Wid_NLO_R}.


   The following shows respectively the formula used for each corresponding  model depicted in the first column of the table      
\begin{align}
W^{2} &=W^2_{\rm{NG_{(\ell o)}}}+W^2_{\rm{PK_{(\ell o)}}},\label{Wid_LO_R}\\
W^{2} &=W^2_{\rm{NG_{(\ell o)}}}+W^2_{\rm{NG_{(n\ell o)}}}+W^2_{\rm{PK_{(\ell o)}}},\label{Wid_NLO_R}\\
W^{2} &=W^2_{\rm{NG_{(\ell o)}}}+W^2_{\rm{PK_{(\ell o)}}}+W^2_{\rm{PK_{(n\ell o)}}},\label{Wid_NLO_R_NLO}
\end{align}

    The resultant fits to the smooth string consisting  of the rigidity terms Eqs.~\eqref{Wid_LO_R} added to the next-to-leading order solution of the NG action Eq.~\eqref{WidthNLO} are inlisted in Tables.~\ref{Fits1NGR} and ~\ref{Fits2NGR}.

    The values of $\chi^{2}$ in the two Tables.~\ref{Fits1NGR} and ~\ref{Fits2NGR} are indicating significant reduction in the values of $\chi^{2}$ at the temperature $T/T_c=0.9$ compared to returned residuals (Table~\ref{LO&NLO_fits_PlanesXT09}) considering only the NG string Eq.~\eqref{WidthNLO}. Moreover, the fit to stiff string are returning good values of $\chi^{2}$ on the whole the intermediate source separation distances at all transverse planes along the tube.    

\begin{table*}[!hpt]
	\begin{center}
		\begin{ruledtabular}
			\begin{tabular}{ll|cccc|cccc|cccc}
				\multirow{2}{*}{$T/T_c=0.9$} & 
				&\multicolumn{4}{c}{$R \in[0.4,0.7]$ fm} 
				&\multicolumn{4}{c}{$R \in[0.5,0.8]$ fm}
				&\multicolumn{4}{c}{$R \in[0.6,0.9]$ fm}\\
				\multicolumn{2}{c|}{\tiny{String Type}} 
				&\multicolumn{1}{c}{$\chi^{2}$} &\multicolumn{1}{c}{$R_0$} 
				&\multicolumn{1}{c}{$b_2$} &\multicolumn{1}{c}{$b_4$}&\multicolumn{1}{c}{$\chi^2$}&\multicolumn{1}{c}{$R_0$}&\multicolumn{1}{c}{$b_2$}&\multicolumn{1}{c}{$b_4$} &\multicolumn{1}{c}{$\chi^2$}&\multicolumn{1}{c}{$R_0$}&\multicolumn{1}{c}{$b_2$}&\multicolumn{1}{c}{$b_4$}\\\hline			
                                \multirow{2}{*}{\begin{turn}{0} {\scriptsize{$z=1$}} \end{turn}}	 
                                 &\small{$NG_{{\ell o}}$}&2.25&10.9(2)&36(2.8)&-23(1.9)&2.01&6(2.0)&120(34)&-97(31.0)&0.001&17(5.8)&-95(79)&103(74)\\
                                \multicolumn{1}{c}{} &\small{$NG_{{n\ell o}}$}&0.23&12.3(3)&17(2.5)&-15(1.8)&0.13&10(2.9)&45(44)&-39(36.8)&0.02&6(5.7)&19(74)&-14(63)\\\hline                                
\multirow{2}{*}{\begin{turn}{0} {\scriptsize{$z=2$}} \end{turn}}	 
 &\small{$NG_{{\ell o}}$}&6.76&4.9(2)&35(2.4)&-23(1.6)&4.04&-1(2.0)&150(30)&-126(26)&0.06&18(5.1)&-102(68)&113(64)\\
\multicolumn{1}{c}{} &\small{$NG_{{n\ell o}}$}&2.82&4.8(2)&26(2.1)&-19(1.6)&1.33&-0.4(2.4)&110(37)&-89(31)&0.18&7(5)&13(64)&-6.(54)\\\hline
\multirow{2}{*}{\begin{turn}{0} {\scriptsize{$z=3$}} \end{turn}}	 
 &\small{$NG_{{\ell o}}$} &9.93&1.3(2)&29(2.4)&-24(1.6)&7.71&-6.4(1.7)&175(30.5)&-153(27)&0.33&20(5.4)&-156(73.5)&160(68)\\
\multicolumn{1}{c}{} &\small{$NG_{{n\ell o}}$}&5.9&0.1(2)&27(2.1)&-22(1.6)&4.85&-8(2.5)&157(39)&-130(32)&0.33&9(5.3)&-41(68.4)&40(58)\\\hline
\multirow{2}{*}{\begin{turn}{0} {\scriptsize{$z=4$}} \end{turn}}	 
 &\small{$NG_{{\ell o}}$} &--- &--- &--- &---&9.55&-6.9(1.7)&124.5(30.1)&-118.4(27)&0.50&22.9(6.1)&-250.8(80)&237(75)\\
\multicolumn{1}{c}{} &\small{$NG_{{n\ell o}}$}&---&---&---&---&6.65&-8.5(2.5)&108.8(38)&-96.3(32)&0.70 &11.9(5.9)&-133.9(75)&112.1(63) \\\hline
\multirow{2}{*}{\begin{turn}{0} {\scriptsize{$z=\frac{R}{2}$}} \end{turn}}	 
 &\small{$NG_{{\ell o}}$}&13.57&2.3(2)&11.3(2.4)&-8.6(1.6)&2.52&-6(1.7)&157(31)&-138(27.4)
&0.003&33(6)&-261(81)&267(76)\\
\multicolumn{1}{c}{} &\small{$NG_{{n\ell o}}$}&8.49&1.3(2)&8(2.1)&-6(1.5)&1.01&-7(2.5)&140(39)&-116(32)&0.01&21(5.9)&-135(75)&123(64)\\             
			\end{tabular}		
		\end{ruledtabular}
	\end{center}
	\caption{ The returned $\chi^2$ from the fits of either the leading order NG string width Eq.~\eqref{WidthLO} or next-to-leading width of NG string Eq.~\eqref{WidthNLO} together with terms due to two terms in the boundary action $W^2_{b_2}+W^2_{b_4}$ given by models Eq.~\eqref{Wid_LO_Boundb2b4} and Eq.~\eqref{Wid_NLO_Boundb2b4}. The Monte-Carlo lattice data of the square-width of the action density width.}
\label{Fits3NGb2b4}	
\end{table*}  

\begin{table*}[!hpt]
	\begin{center}
		\begin{ruledtabular}
			\begin{tabular}{ll|cccc|cccc|cccc}
				\multirow{2}{*}{$T/T_c=0.9$} &  
				&\multicolumn{4}{c}{$R \in[0.4,1.2]$ fm} 
				&\multicolumn{4}{c}{$R \in[0.5,1.2]$ fm}
				&\multicolumn{4}{c}{$R \in[0.6,12]$ fm}\\
				\multicolumn{2}{c|}{\scriptsize{~~~String Type}} 
				&\multicolumn{1}{c}{$\chi^{2}$} &\multicolumn{1}{c}{$R_0$} &\multicolumn{1}{c}{$b_2$} &\multicolumn{1}{c}{$b_4$} 
				&\multicolumn{1}{c}{$\chi^2$} &\multicolumn{1}{c}{$R_0$} &\multicolumn{1}{c}{$b_2$}&\multicolumn{1}{c}{$b_4$} 
				&\multicolumn{1}{c}{$\chi^2$} &\multicolumn{1}{c}{$R_0$} &\multicolumn{1}{c}{$b_2$}&\multicolumn{1}{c}{$b_4$} \\
				\hline 
\multirow{2}{*}{\begin{turn}{0} {\scriptsize{$z=1$}} \end{turn}}	 
 &\small{$NG_{{\ell o}}$}&52.3&11.6(2)&45(2.4)&-29(1.6)&52.2&12(1.7)&33(31)&-18(27.8)&0.70&25(2.5)&-121(38.0)&131(34.8)\\
\multicolumn{1}{c}{} &\small{$NG_{{n\ell o}}$}&6.38&12.6(2)&18(2)&-16(1.6)&5.66&14(2.2)&-11(34.5)&7(28.7)&1.43&17(2.5)&-36(36.7)&31(30.8)\\\hline
\multirow{2}{*}{\begin{turn}{0} {\scriptsize{$z=2$}} \end{turn}}	 
 &\small{$NG_{{\ell o}}$} &92.9 &5.8(1)&44(2.0)&-30(1.4)&92.5&4.8(5)&62(27.2)&-45(24.3)&1.5&
22(2.4)&-153(35)&161(32.5)\\
\multicolumn{1}{c}{} &\small{$NG_{{n\ell o}}$}&28.2&5.3(2)&30(1.8)&-22(1.3)&27.5&7(1.9)&4(30.8)&-0.5(25.7)&3.08&
14(2.4)&-67(34.1)&63(28.7)\\\hline
\multirow{2}{*}{\begin{turn}{0} {\scriptsize{$z=3$}} \end{turn}}	 
 &\small{$NG_{{\ell o}}$}&274.497&2.7(2)&44(2.0)&-34(1.4)&245.0&10(1.4)&-87(24.4)&82(21.6)&4.06& 25(1.7)&-227(26.0)&226(23.6) \\
\multicolumn{1}{c}{} &\small{$NG_{{n\ell o}}$}&133.4&1.3(2)&36(1.8)&-28(1.4)&95.76&10(1.6)&-117(25.2)&99.5(21.0)&5.58&17(1.6)&-144(25.3)&128(21)\\\hline
\multirow{2}{*}{\begin{turn}{0} {\scriptsize{$z=4$}} \end{turn}}	 
 &\small{$NG_{{\ell o}}$} &---&---&---&---&190.4&5(1.4)&-61(26.1)&49(23.2)&3.49&24(2.0)&-270(30.2)&255(27.6)\\
\multicolumn{1}{c}{} &\small{$NG_{{n\ell o}}$}&---&---&---&---&84.57&6(1.8)&-111(28.0)&87(23.3)&3.99&16(2.1)&-186(29.3)&157(24.6)\\\hline
\multirow{2}{*}{\begin{turn}{0} {\scriptsize{$z=\frac{R}{2}$}} \end{turn}}	 
 &\small{$NG_{{\ell o}}$}&40.36&2.8(2)&17(2.1)&-12(1.5)&24.20&-4(1.7)&137(29.9)&-119(26.6)&0.29&12(3.6)&-59(49.9)&67(46.3)\\
\multicolumn{1}{c}{} &\small{$NG_{{n\ell o}}$}&14.52&1.7(2)&10(1.9)&-7(1.4)&6.96&-4(2.3)&108(35.7)&-89(29.7)&0.07&3(3.6)&25(47.5)&-17(40.3)\\
			\end{tabular}		
		\end{ruledtabular}
	\end{center}
	\caption{Same as Table~\ref{Fits3NGb2b4}; however, the fit range includes larger color source separations.}
        \label{Fits4NGb2b4}
\end{table*}
\begin{table*}[!hpt]
	\begin{center}
		\begin{ruledtabular}
			\begin{tabular}{ll|cccc|cccc}
				\multirow{2}{*}{$T/T_c=0.9$} &
				&\multicolumn{4}{c}{$R \in[0.4,1.2]$ fm} 
				&\multicolumn{4}{c}{$R \in[0.5,1.2]$ fm}\\
				\multicolumn{2}{c|}{\scriptsize{String Type}} 
				&\multicolumn{1}{c}{$\chi^{2}$} &\multicolumn{1}{c}{$R_0$}&\multicolumn{1}{c}{$\alpha$}&\multicolumn{1}{c}{$b_2$} 
                                &\multicolumn{1}{c}{$\chi^2$}&\multicolumn{1}{c}{$R_0$}&\multicolumn{1}{c}{$\alpha$}&\multicolumn{1}{c}{$b_2$}\\\hline 
                                \multirow{2}{*}{\begin{turn}{0} {\scriptsize{$z=1$}} \end{turn}}	 
                                &\small{$NG_{{\ell o}}$}&3.90&30(1.1)&3.1(2)&7.2(2)&0.0&3.54&31(2.5)&3.3(5)\\
                                \multicolumn{1}{c}{} &\small{$NG_{{n\ell o}}$}&3.62&21.3(9)&1.2(1)&-11.4(7)&
1.70&18(2.3)&0.7(4)&-7(2.9)\\\hline                                    
\multirow{2}{*}{\begin{turn}{0} {\scriptsize{$z=2$}} \end{turn}}	 
 &\small{$NG_{{\ell o}}$}&10.72&25.6(9)&3.3(1)&4.9(2)&7.89&29(2.3)&3.9(4)&3.4(8) \\
\multicolumn{1}{c}{} &\small{$NG_{{n\ell o}}$}&4.87&17.6(8)&1.63(9)&-10.3(6)&4.62&17(2.0)&1.5(3)&-9(2.6)\\\hline
\multirow{2}{*}{\begin{turn}{0} {\scriptsize{$z=3$}} \end{turn}}	 
 &\small{$NG_{{\ell o}}$}& 60.67 & 26.3(9) & 3.9(1) &  -0.8(2) & 31.51 & 36(2)& 5.9(4)&-5.5(9)\\
\multicolumn{1}{c}{} &\small{$NG_{{n\ell o}}$}& 23.34 & 17.6(8)& 2.16(9)&-16.1(6)&16.02&22(1.8)&2.9(3)&-22(2) \\\hline
\multirow{2}{*}{\begin{turn}{0} {\scriptsize{$z=4$}} \end{turn}}	 
 &\small{$NG_{{\ell o}}$}& --  &-- &--&--& 32.63 &32(2)&5.6(4)&-15.9(9) \\
\multicolumn{1}{c}{} &\small{$NG_{{n\ell o}}$}& -- &-- & --& --&18.0 &18.66 & 2.9(3)&-30(2.7)\\\hline
\multirow{2}{*}{\begin{turn}{0} {\scriptsize{$z=\frac{R}{2}$}} \end{turn}}	 
 &\small{$NG_{{\ell o}}$}&0.84&27(5.1)&4.0(8)&-29(9.3)&0.1&20(7.8)&3.2(9)&20(52.0)\\
&\small{$NG_{{n\ell o}}$}&1.15&17(4.6)&2.0(6)&-37(13.8)&$1.7 \times 10^{-7}$ &5(6.9)&1.1(6)&57(57.5)\\
			\end{tabular}		
		\end{ruledtabular}
	\end{center}
	\caption{The returned $\chi^2$ from the fits of mean-square width of flux-tube to the LO and the NLO width of NG string combined with leading contribution of rigidity and boundary terms $W^{2}_{PK}+W^{2}_{b_2}$. The two raws entries in each cell corresponds to the models given Eq.~\eqref{Wid_LO_Rigid_Boundb2} and Eq.~\eqref{Wid_NLO_Rigid_Boundb2}.}
\label{Fits5NGb2R}	
\end{table*}  

\begin{table*}[!hpt]
	\begin{center}
		\begin{ruledtabular}
			\begin{tabular}{ll|ccccc|ccccc}
				\multirow{2}{*}{$T/T_c=0.9$} &\scriptsize{P/I} 
				&\multicolumn{5}{c}{$R \in[0.4,1.2]$ fm} 
				&\multicolumn{5}{c}{$R \in[0.5,1.2]$ fm}\\
				\multicolumn{2}{c|}{\scriptsize{String Type}} 
				&\multicolumn{1}{c}{$\chi^{2}$} &\multicolumn{1}{c}{$R_0$}&\multicolumn{1}{c}{$\alpha$}&\multicolumn{1}{c}{$b_2$} &\multicolumn{1}{c}{$b_4$}
                                &\multicolumn{1}{c}{$\chi^2$}&\multicolumn{1}{c}{$R_0$}&\multicolumn{1}{c}{$\alpha$}&\multicolumn{1}{c}{$b_2$}&\multicolumn{1}{c}{$b_4$}\\\hline
                                \multirow{2}{*}{\begin{turn}{0} {\scriptsize{$z=1$}} \end{turn}}	 
                                &\small{$NG_{{\ell o}}$}&3.16&32(2.9)&3.5(5)&1(6.7)&4(5.0)&1.11&36(3.7)&3.6(5)&-45(33.2)&45(29)\\
                                \multicolumn{1}{c}{} &\small{$NG_{{n\ell o}}$}&1.92&18(2.6)&0.7(3)&0.2(8.9)&-6(5.0)&1.57&19(3.2)&0.7(3)&-20(34.9)&10(28.8)\\\hline
\multirow{2}{*}{\begin{turn}{0} {\scriptsize{$z=2$}} \end{turn}}	 
 &\small{$NG_{{\ell o}}$}&6.78&30(2.7)&4.1(4)&-6(5.8)&9(4.4)&4.12&34(3.5)&4.3(5)&-54(29.9)&51(26.3)\\
\multicolumn{1}{c}{} &\small{$NG_{{n\ell o}}$}&4.73 &17(2.3)&1.5(3)&-7(7.9)&-2(4.4)&4.18&18(3)&1.5(3)&-30(31.6)&17(26)\\\hline
\multirow{2}{*}{\begin{turn}{0} {\scriptsize{$z=3$}} \end{turn}}	 
&\small{$NG_{{\ell o}}$}&21.88&39(2.3)&6.2(4)&-34(5.4)&25(4.0)&10.28&43(2.6)&6.0(4)&-115(24.5)&98(21.7)\\
\multicolumn{1}{c}{} &\small{$NG_{{n\ell o}}$}& 3.35 &23(2) &3.0(3) &-38(7)&12(4)&8.76 & 25(2)&2.7(3)&-90(25)& 57(21) \\\hline
\multirow{2}{*}{\begin{turn}{0} {\scriptsize{$z=4$}} \end{turn}}	 
 &\small{$NG_{{\ell o}}$}& --&--&--&--&--&7.0&42(3)&6.1(4)& -152(27)& 120(24)\\
\multicolumn{1}{c}{} &\small{$NG_{{n\ell o}}$}& -- &-- &-- &--&--&5.76&24(3)&2.8(3)& -127(28)&81(23)\\\hline
\multirow{2}{*}{\begin{turn}{0} {\scriptsize{$z=\frac{R}{2}$}} \end{turn}}	 
 &\small{$NG_{{\ell o}}$}&1.65&24(3.4)&3.6(6)&-25(7.0)&19(5.3)&0.14&20(5.1)&3(6)&21(38)&-21(33.3)\\
\multicolumn{1}{c}{} &\small{$NG_{{n\ell o}}$}&3.67&12(3.1)&1.3(4)&-22(9.7)&10(5.3)&0.04&6(4.5)&1.1(4)&55(40.9)&-52(32.8)\\
\end{tabular}		
\end{ruledtabular}
\end{center}
        \caption{Same as Table.\ref{Fits5NGb2R}, however, the next to leading contribution from the boundary action has been considered $W^{2}_{b_4}$. The fits are for the model given by Eq.~\eqref{Wid_LO_Rigid_Boundb2b4} and Eq.~\eqref{Wid_NLO_Rigid_Boundb2b4}.}
\label{Fits6NGb2b4R}	
\end{table*}  

\begin{table*}[!hpt]
	\begin{center}
		\begin{ruledtabular}
			\begin{tabular}{ll|cccc|cccc|cccc}
				\multirow{2}{*}{$T/T_c=0.8$} & 
				&\multicolumn{4}{c}{$R \in[0.4,1.2]$ fm} 
				&\multicolumn{4}{c}{$R \in[0.5,1.2]$ fm}
				&\multicolumn{4}{c}{$R \in[0.6,12]$ fm}\\
				\multicolumn{2}{c|}{\scriptsize{~~~String Type}} 
				&\multicolumn{1}{c}{$\chi^{2}$} &\multicolumn{1}{c}{$R_0$} &\multicolumn{1}{c}{$b_2$} &\multicolumn{1}{c}{$b_4$} 
				&\multicolumn{1}{c}{$\chi^2$} &\multicolumn{1}{c}{$R_0$} &\multicolumn{1}{c}{$b_2$}&\multicolumn{1}{c}{$b_4$} 
				&\multicolumn{1}{c}{$\chi^2$} &\multicolumn{1}{c}{$R_0$} &\multicolumn{1}{c}{$b_2$}&\multicolumn{1}{c}{$b_4$} \\
				\hline 
\multirow{2}{*}{\begin{turn}{0} {\scriptsize{$z=2$}} \end{turn}}	 
 &\small{$NG_{{\ell o}}$} &92.9 &5.8(1)&44(2.0)&-30(1.4)&92.5&4.8(5)&62(27.2)&-45(24.3)&1.5&
22(2.4)&-153(35)&161(32.5)\\
\multicolumn{1}{c}{} &\small{$NG_{{n\ell o}}$}&28.2&5.3(2)&30(1.8)&-22(1.3)&27.5&7(1.9)&4(30.8)&-0.5(25.7)&3.08&
14(2.4)&-67(34.1)&63(28.7)\\\hline
\multirow{2}{*}{\begin{turn}{0} {\scriptsize{$z=3$}} \end{turn}}	 
 &\small{$NG_{{\ell o}}$}&274.497&2.7(2)&44(2.0)&-34(1.4)&245.0&10(1.4)&-87(24.4)&82(21.6)&4.06& 25(1.7)&-227(26.0)&226(23.6) \\
\multicolumn{1}{c}{} &\small{$NG_{{n\ell o}}$}&133.4&1.3(2)&36(1.8)&-28(1.4)&95.76&10(1.6)&-117(25.2)&99.5(21.0)&5.58&17(1.6)&-144(25.3)&128(21)\\\hline
\multirow{2}{*}{\begin{turn}{0} {\scriptsize{$z=4$}} \end{turn}}	 
 &\small{$NG_{{\ell o}}$} &---&---&---&---&190.4&5(1.4)&-61(26.1)&49(23.2)&3.49&24(2.0)&-270(30.2)&255(27.6)\\
\multicolumn{1}{c}{} &\small{$NG_{{n\ell o}}$}&---&---&---&---&84.57&6(1.8)&-111(28.0)&87(23.3)&3.99&16(2.1)&-186(29.3)&157(24.6)\\
			\end{tabular}		
		\end{ruledtabular}
	\end{center}
	\caption{Same as Table~\ref{Fits6NGb2b4R}; however, the temperature is lowered to $T/T_{c}=0.8$.}
       \label{Fits7NGb2b4RT08}	
\end{table*}


   The solid and dashed lines in the plot of Figs.~\ref{MidPlane} and ~\ref{FitsPlanes} corresponding to the NG string in the interaction approximations Eqs.\eqref{WidthNLO} and rigid strings Eqs.~\eqref{Wid_LO_R} and ~\eqref{Wid_NLO_R} show the dramatic improvement in the fits with respect to the stiff strings when considering fit range covering the whole fit range $R \in [0.5-1.2]$ fm.

   The $\chi^2$ values returned from the fit to a string model~\eqref{Wid_LO_R} at only the LO perturbation from both NG and PK string, indicate improvement as well compared to the free string NG string. However, We find that considering both corrections returns improved values of $\chi^{2}_{dof}=11.09/6$ over the shorter source separation interval $R\in [0.5, 1.2]$ fm.

   Nevertheless, drawing a comparison between the fit behavior of any of the leading and the formula for NLO in the extrinsic curvature Eq.~\eqref{Wid_NLO_R_NLO} reveals a subtle difference in the returned $\chi^{2}$ on fit interval $R \in [0.5, 1.2]$ fm the corrections within the uncertainties of the measurements. Effects of the NLO in perturbation Eq.~\eqref{Wid_NLO_R_NLO} could be relevant when considering smaller distances, finer lattices or much higher resolutions in general.

     The improvement in the fit with respect to the rigid strings compared to that obtained merely on the basis of pure NG string is  displayed in Figs.~\ref{MidPlane}. For source separation range $R \in [0.5, 0.7]$ fm, the rendering in Fig.~\ref{MidPlane} of the fitted width of the pure NG string at either LO or NLO corrections exhibits significant deviations from the data compared to the corresponding fits in Fig.~\ref{MT09} over the fit interval $R \in [0.7, 1.2]$ fm. The plots in Fig.~\ref{MidPlane} indicate the incompetence of the pure NG string as a physical description integrating out the properties of the QCD flux tubes.

   Apart from mitigate deviation at short distances when describing the flux-tubes width over planes other than the middle, .i.e,  $x=2,3,4$ planes (Table~\ref{Fits1NGR},\ref{Fits2NGR}). Similar assertion still holds true that the match is enhanced with respect to the rigid string models over larger intermediate separation distances as evidently displayed in Fig.~\ref{FitsPlanesR}.

   The fit to the NG approximation Eq.~\eqref{Wid_NLO_R} returns good values of $\chi^2$ for the mean square width of the string in the middle plane. However, the fit to the stiff string Eq.~\eqref{Wid_LO_R} exhibits improvements with respect the planes near to the color sources, this matches the intuitive picture that rigidity effects/resistance to bending may be more stringent near the string ends.

\subsection{L\"uscher-Wiesz string with two boundary-terms in the action}
     
  The corrections provided by the boundary action to static $Q\bar{Q}$ potential seem to explain to some extend the deviations appearing when constructing the static mesonic states with Polyakov loop correlators~\cite{Bakry:2017utr}. At a higher temperature, the inclusion of the boundary corrections up to the fourth order $b_4$ together with string rigidity the $Q\bar{Q}$ potential has been found to be viable~\cite{Bakry:2017utr} is well described in providing good fits for distances as small as $R=0.5$ fm.

    The goal in this section is to compare the analytic estimate of the mean-square width resulting from the boundary terms in Lüscher-Weisz (LW) effective string action Eq.~\eqref{GBaction}. This could be compatible with the energy fields set up by a static mesonic configurations. We consider the perturbative expansion of two boundary terms at the order of fourth and six derivative given by Eq.~\eqref{b2Complete} and Eq.~\eqref{b4Complete} respectively.
    
    In the following we select possibly interesting combinations of boundary terms with LO and NLO Nambu-Goto and rigid string with Eqs.~\eqref{1wb2} and ~\eqref{wb4}. We consider 

\begin{align}  
  W^{2}&=W^{2}_{{NG_{(\ell o)}}}+W^{2}_{b_2},\label{Wid_LO_Boundb2}\\
  W^{2}&=W^{2}_{{NG_{(\ell o)}}}+W^{2}_{{NG_{(n\ell o)}}}+W^{2}_{b_2}, \label{Wid_NLO_Boundb2}\\
  W^{2}&=W^{2}_{{NG_{(\ell o)}}}+W^{2}_{b_2}+W^{2}_{b_4},           \label{Wid_LO_Boundb2b4}\\
  W^{2}&=W^{2}_{{NG_{(\ell o)}}}+W^{2}_{{NG_{(n\ell o)}}}+W^{2}_{b_2}+W^{2}_{b_4},  \label{Wid_NLO_Boundb2b4}
\end{align}
which excludes the rigidity structure of the string. Similar to the foregoing sections, comparison with the broadening of the width at each selected transverse plane can be drawn with that of the corresponding width of various models string models. We summarize the observations on the resultant $\chi^{2}$ and fit parameters in Tables.~\ref{Fits1NGb2},~\ref{Fits2NGb2} and \ref{Fits3NGb2b4}.

\begin{itemize}  

\item Tables~\ref{Fits1NGb2} and ~\ref{Fits2NGb2}\\
  In the first two column are the returned parameter is the UV cutoff $R_{0}$ and the following entries are for the boundary action couplings $b_2$.

  The first observation is that the fit over short distance ranges in Table~\ref{Fits1NGb2} shows improvements when including the first boundary correction terms as in model Eq.~\eqref{Wid_LO_Boundb2} compared to the corresponding LO NG model. The values f $\chi^{2}$ is further reduced upon switching on the string's self-interaction Eq.~\eqref{Wid_NLO_Boundb2} together with the boundary corrections. 

  Secondly, one can remark that the LO width of NG with $W_{b_2}^{2}$ correction model of Eq.~\eqref{Wid_LO_Boundb2} is providing smaller values of residual than the NLO pure NG string.
  
\item Tables~\ref{Fits3NGb2b4} and ~\ref{Fits4NGb2b4}\\
  The first Table.~\ref{Fits3NGb2b4} covers fit results over selected short intervals. The immediate observation that the consideration of the next six derivative term in the boundary action had improved the fits over these intervals in particular for distances commencing from $R=0.4$ fm, even without considering the NLO term from NG string.
    
  For longer distances (Table.~\ref{Fits4NGb2b4}) the inclusion of the NLO term from NG string seems to be in effect, nevertheless, the fits at the planes $z=2, z=3$ are still show sensible deviations from the lattice data.
 
\end{itemize}

   Figure~\ref{BoundFitMid}-(a) illustrates the resultant fitted curves of the NLO width at the middle plane of NG string together with leading boundary term $W^{2}_{b_2}$ Eq.~\eqref{Wid_NLO_Boundb2}. The fits of the string model which employs the next nonvanshing boundary term $W^{2}_{b_4}$ Eq~\eqref{Wid_NLO_Boundb2b4}, is depicted in Fig~\ref{BoundFitMid}-(b). Here we evidently see the match with the LGT data up to a surprisingly small distances $R=0.4$ fm.

   The plot in Fig.~\ref{BoundFitPlanes} shows the contribution of the boundary action to the width profile at two consequtive planes from the quark, namely, $z=1$ and $z=2$. The line are for the fits of the string model with two boundary terms $(b_2,b_4)$ Eq.~\eqref{Wid_NLO_Boundb2b4} over interval $R \in [0.5,1.2]$ fm. For the planes apart from the middle plane it seems the mismatch is
evident at the large source separation distances.  
   
  More variants of the string models can be attained by switching on the rigid properties of the QCD string model,

 \begin{align} 
  W^{2}&=W^{2}_{{\rm{NG}_{(\ell o)}}}+W^2_{\rm{PK_{(\ell o)}}}+W^{2}_{b_2},\label{Wid_LO_Rigid_Boundb2}\\
  W^{2}&=W^{2}_{{\rm{NG}_{(\ell o)}}}+W^{2}_{{\rm{NG}_{(n\ell o)}}}+W^2_{\rm{PK_{(\ell o)}}}+W^{2}_{b_2},\label{Wid_NLO_Rigid_Boundb2}\\  
\end{align} 
 in the above only the first boundary correction $W^{2}_{b_2}$ is encompassed.

  Table.~\ref{Fits5NGb2R} enlists the returned fit parameters and the residual. Both models of Eq.~\eqref{Wid_LO_Rigid_Boundb2} and Eq.~\eqref{Wid_NLO_Rigid_Boundb2} have a very good match with the numerical data at all planes up to source separation distances as small as $R=0.4$ fm.

  However, subtle differences in the values of $\chi^2$ are observed when considering the next-surviving boundary term $W^{2}_{b_4}$,
  
\begin{align}
W^{2}&=W^{2}_{{\rm{NG}_{(\ell o)}}}+W^2_{\rm{PK_{(\ell o)}}}+W^{2}_{b_2}+W^{2}_{b_4},\label{Wid_LO_Rigid_Boundb2b4}\\
W^{2}&=W^{2}_{{\rm{NG}_{(\ell o)}}}++W^{2}_{{\rm{NG}_{(n\ell o)}}}+W^2_{\rm{PK_{(\ell o)}}}+W^{2}_{b_2}+W^{2}_{b_4},\label{Wid_NLO_Rigid_Boundb2b4}
\end{align}  

in the models of Eq.~\eqref{Wid_LO_Rigid_Boundb2} and Eq.\eqref{Wid_NLO_Rigid_Boundb2} as the inspection of Table.~\ref{Fits6NGb2b4R} divulges. The good fit, nevertheless,  is remarkable in these model at planes from $z=3$ and$z=4$ from the quarks. This seems to
suggest restoring forces, by virtue of the inclusion of the rigidity, that pull down the strings into its classical configuration. 


  The expectations of an almost flat width geometry along the transverse planes is consistent with the analysis shown in Figs.~\ref{DIFFT09}, which indicates that the thermal effects strongly diminishes near the QCD plateau region~\cite{PhysRevD.85.077501}. In Figs.~\ref{DIFFT08} the render of the action densities corresponding to the both temperatures $T/T_{c}=0.8$ and $T/T_{c}=0.9$ unveils an independent prolate-shaped action density for the color map.
 
 These are two typical instances where the string's width profile is exhibiting a constant width along the tube. The first is due to the diminish of  thermal effects near the end of QCD plateau, the second manifests at the intermediate color source separations and the temperature close to the deconfinement point as a result of the role played by the string-self interactions. This is culminated in the squeeze/suppression along the transverse planes. Figs.~\ref{DIFFT09} and ~\ref{DIFFT08} disclose the fact that the geometry of the density isolines are quite independent from both the width profile

\section{Summary of numerical results}

  The corrections received from the Nambu-Goto (NG) action expanded up next to leading order terms have been set into comparison with the corresponding  $SU(3)$ Yang-Mills lattice data in four dimensions. The region under scrutiny is where the free string picture poorly describe the energy profile. The considered source separation are $R=0.5$ to $R=1.2$ fm for two temperatures scales near the end of QCD plateau and just before the critical point $T/T_{c}=0.8$ and $T/T_{c}=0.9$.

   The theoretical predictions laid down by both the LO and the NLO approximations of Nambu-Goto string show a good fit behavior for the data corresponding to the $Q\bar{Q}$ potential near the end of the QCD plateau region at $T/T_{c}=0.8$. The fit returns almost the same parameterization behavior with negligible differences for the measured zero temperature string tension $\sigma_{0}a^{2}$. The returned value of this fit parameter is in agreement with the measurements at zero temperature~\cite{Koma:2017hcm}
  
   On the otherhand, at a higher temperature near the deconfinement point $T/T_{c}=0.9$, the values of $\chi^{2}$ indicate improvements with respect to the fits to the NLO width profile of the pure NG string Eq.(3.11) compared to the leading-order approximation. Nevertheless, the NLO approximation does not provide an accurate match with the numerical data except at R > 0.8 fm. The fits of the $Q\bar{Q}$ potential data to the Nambu-Goto string model considering either of its approximation schemes of NG string return large values of $\chi^{2}$ if the fit region span the whole source separation distances $R=0.5$ fm to $R=1.2$ fm. The fits at next to leading order approximation of the NG string show some improvements on each corresponding fit interval. In general, the values of the residuals decrease by the exclusion of the data points at short distances for both approximations.

   The inclusion of leading boundary term of L\"uscher-Weisz action $W_{b_2}^2$ in the approximation scheme improves the fits at all the considered source separations. The fits of the string model employing the next non-vanshing boundary term $W^{2}_{b_4}$ Eq~\eqref{Wid_NLO_Boundb2b4} displays evident match with the LGT data up to a surprisingly small distances $R=0.4$ fm. Although large distance deviations reappear for planes away from the middle plane of the flux-tube, we see that a good match is recovered when considering rigid properties of the string.  

    We found that the rigid string width profile accurately matches the width measured from the numerical lattice data near the deconfinement point. This suggests that the rigidity effects can be very relevant to the correct description of Monte-Carlo data of the field density and motivates scrutinizing the stiffness physics of QCD-flux-tube in other frameworks~\cite{Cea:2014uja}.

   For the considered fit interval $R \in [0.5, 1.2]$ fm the corrections received from the two-loop in the extrinsic curvature is vey small and within the uncertainities of the measurements. The next to leading-order in perturbation Eq.(3.24) could be be very relevant when considering smaller distances, finer lattices, or other gauge models.

   Although the rigidity effects seems to be very similar to considering higher-order terms in NG string, we find that considering both two terms return improved values of $\chi^2_{dof}=10.58/6$ for shorter distances $R \in [0.5, 1.2]$ fm. Drawing a comparison between the fit behavior of both the LO and the NLO formula for the extrinsic curvature Eq.(2.19) and Eq.(3.24) reveals a subtle difference in the  returned $\chi^{2}$ and around (20 to 30) percent change in the value of the rigidity parameter.
    
   At higher temperature $T/T_c=0.9$, the color tube exhibits a suppressed growth profile in the intermediate region. The fits considering both intermediate and asymptotic color source separation distances show noticeable improvement with respect to the string self-interacting picture (NLO) compared to that obtained on the basis of the free string approximation. Nevertheless, the next to leading approximation does not provide an accurate match the numerical data. This manifests as significantly large values of the returned $\chi^{2}$ when considering distances less than $R\textless 0.8$ fm.
       
  The oscillations of a free NG string fixed at the ends by Dirichlet boundaries traces out a nonuniform width profile with a geometrical curved fine structure. This is detectable~\cite{PhysRevD.82.094503} at source separations $R>1.0$ fm and near to the critical temperature. However, in the intermediate region the lattice data are not in consistency with the curved width of the free fluctuating string. The fits to mean-square width extracted from the NLO expansion of NG string, however, indicate that self interactions flatten the width profile in the intermediate region. The string's self-interactions accounts for the constant width along consecutive transverse plane to the tube in addition to the decrease in slop of the suppressed width broadening. 

  At the end of the QCD plateau region at temperature $T/T_c=0.8$ the constant width property is manifesting at all source separation distances and is in consistency with the pure NG action. These results indicate not only the fade out of the thermal effects at this temperature but also indicate a form of the action density map independent from the geometrical changes induced by the temperature. That is, the main features of the density map would persist at lower and zero temperature.
  
\section{Conclusion}   
   In this investigation, we discussed the effective bosonic string model of confinement in the vicinity of critical phase transition point~\cite{Bakry:2018kpn}. We conclude that the free Nambu-Goto string can be a good description to energy profile of QCD flux-tubes up to temperatures on the QCD plateau. With the gradual decrease of the string tension the pure NG string does not precisely describe the lattice Mont-Carlo data even, at two-loop orders. Nevertheless, we evidence that the effective bosonic string model is competently integrating out the physical properties of the flux-tube when including symmetry effects of the boundary action and rigidity properties into its paradigm.


\begin{acknowledgments}
  We thank S. Brandt, M. Casselle for their useful comments and discussions. This work has been funded by the Chinese Academy of Sciences President's International Fellowship Initiative grants No.2015PM062 and No.2016PM043, the Polish National Science Centre (NCN) grant 2016/23/B/ST2/00692, the Recruitment Program of Foreign Experts, NSFC grants (Nos.~11035006,~11175215,~11175220) and the Hundred Talent Program of the Chinese Academy of Sciences (Y101020BR0).
\end{acknowledgments}

\bibliography{Biblio2}%

\end{document}